\documentclass[12pt]{book}
\usepackage{amsfonts}
\usepackage{amsmath}
\usepackage{amssymb}
\usepackage{graphicx}
\usepackage[all]{xy}
\newcommand{\PL}[1]{{\cal PL}(#1)} 
\newcommand{\R}{\cal R} 
\newcommand{\C}{\cal C} 
\newcommand{\Ln}[1]{{\cal L}(#1)}  
\newcommand{\cA}{\cal A}                
\newcommand{\cB}{\cal B}                

\newcommand{\A}{{\hat A}}

\newcommand{\E}{{\hat E}}
\renewcommand{\P}{{\hat P}}
\renewcommand{\S}{{\cal S}}
\newcommand{\Hi}{{\cal H}}

\newcommand{\BH}{\mathcal{B(H)}}
\newcommand{\PH}{\mathcal{P(H)}}
\newcommand{\PV}{\mathcal{P}(V)}

\newcommand{\VH}{\mathcal{V(H)}}
\newcommand{\Su}{\underline S}
\newcommand{\Siu}{\underline{\Sigma}}

\newcommand{\X}{\underline{X}}       

\newcommand{\name}[1]{\ulcorner #1\urcorner}      
\newcommand{\bra}[1]{\langle #1|\,}                            
\newcommand{\ket}[1]{\,|#1\rangle}                             

\newcommand{\LeftDB}{[\mkern-3mu[}  
\newcommand{\RightDB}{]\mkern-3mu]}
\newcommand{\Val}[1]{\LeftDB\,#1\,\RightDB}    
\newcommand{\cop}{{\textbf{Sets}}^{{\C}^{op}}} 
\newcommand{\copv}{{\textbf{Sets}}^{{\cal V(H)}^{op}}} 
\newcommand{\tb}{\textbf} 
\newcommand{\ti}{\textit} 

\newcommand{\goth}{\mathfrak}  

\newcommand{\va}[1]{\tilde{#1}} 
\newcommand{\ra}{\rightarrow} 
\newcommand{\mcl}{\mathcal} 
\newcommand{\aasdf}{$\Omega$} 

\newcommand{\mathC}{\mkern1mu\raise2.2pt\hbox{$\scriptscriptstyle|$}
        {\mkern-7mu\rm C}}            
\newcommand{\mathR}{{\rm I\! R}}      

\newcommand{\mapright}[1]{\smash{
        \mathop{\mbox{\large{$\longrightarrow$}}}\limits^{#1}}}   
     
\newcommand{\mapleft}[1]{\smash{
        \mathop{\mbox{\large{$\longleftarrow$}}}\limits^{#1}}}   

\newcommand\ie{{i.e.},}

\newcommand{\Ga}{\Gamma}

\renewcommand{\l}{\lambda}
\newcommand{\si}{\sigma}
\newcommand{\Si}{\Sigma}
\newcommand{\de}{\delta}
\newcommand{\De}{\Delta}

\renewcommand{\O}{\Omega}
\newcommand{\e}{\epsilon}
\newcommand{\ve}{\varepsilon}
\newcommand{\vr}{\varrho}
\newcommand{\id}{{\rm id}}

\newcommand{\map}{\rightarrow}               

\newcommand{\scl}{Sub_{cl}(\Siu)} 

\newtheorem{definition}{Definition}

\newtheorem{theorem}{Theorem}
\newtheorem{lemma}[theorem]{Lemma}
\newtheorem{remark}[theorem]{Remark}
\newtheorem{axiom}{Axiom}
\newtheorem{proposition}[theorem]{Proposition}

\begin{document}

\title{An Introduction to Topos Physics}
\author{Marios Tsatsos \\
\emph{\small{Submitted in partial fulfillment of the requirements}}\\ 
\emph{\small{for the degree of Masters of Science of the University of London}} 
 }
\date{September 28, 2007}

\maketitle

\tableofcontents

\chapter{Introduction}   
\subsubsection{Why Topos?}

Start talking about topos and physics is not an easy `charge'. While a logician/pure mathematician would feel much confident to lecture on topoi and a theoretical physicist on quantum mechanics, a new `$\underline{~~~~~~~~~}$ist' is needed here to give a thorough and deep explanation of the meanings involved in the present work. And this is mainly because of the complexity and the novelty of the topic. 

However a brief\footnote{since I do not regard myself as a new `$\underline{~~~~~~~~~}$ist'!} introduction of what follows will be attempted, by presenting the motivation and the `ingredients' of it. 

As shown in the title of the present paper, the main issue (however not the only) is that of a topos. A topos is a category, endowed with some further properties. A category is a quite general and abstract mathematical construction. The branch of mathematics that studies the categories and their properties is Category theory, a theory that has come to occupy a central position in mathematics and mathematical physics. Its generality allows many applications in different fields of science; indeed, they can be found in textbooks of logic, algebraic topology, physics \footnote{to be honest, not quite often}, computer science and others. Roughly, it is a general mathematical theory of structures and of systems of structures; its constituents are just `objects' and `arrows' between them. Furthermore, it serves as an alternative to set theory as a foundation for mathematics and, like set theory, it is closely related to logic. But, as will be explained later, it gives rise to logic different than the classical one. Category theory is both an interesting object of philosophical study\footnote{The fact that bears its name from a philosophical term is not an accident.}, and a potentially powerful formal tool for philosophical investigations of concepts such as space, system, and even truth \cite{sep}.
 
 Nonetheless, if one wants to describe the present work in one phrase this would be that: ``We intend to reformulate physics by generalizing the mathematical background, \ie changing the ordinary sets that lie below physics, with structures other than sets. This can be done by employing topos theory to express on it the features of a physical theory'' This structures are studied in the context of category theory. 

 \section{Motivation}
 \subsection{Historical background}
 
Category theory has occurred in the natural route of the abstraction of mathematics. Indeed, the evolution of mathematical thought has the tendency to move on from more ``physical mathematics'' to less concrete ones. In ancient times, the mathematics, developed by Greeks, Arabs and Assyrians were just derived from the necessity of measuring distances and areas; namely geometry. The abstraction of geometrical shapes and `flat spaces` was not far beyond than what could be directly seen and measured. Newton's and Leibniz's  Calculus of the 17th century raised the abstraction in a higher level; extraordinary notions of infinitely small quantities and limits of infinite sequences were introduced. Later on, with the development on curved spaces, logic and number systems, the physical intuition was almost lost; abstract mathematical models and theories are now put in the centre of the mathematical thought.

Especially in the field of logic, the work of the Dutch mathematicians L. Brouwer,  at the beginning of the 20th century and later on by his student A. Heyting, introduced the intuitionism in mathematical logic. Intuitionistic logic contradicts the classical account of truth, which regards a proposition as being always true or false. This type of logic plays a central role in a topos representation of physics, as will become clearer later.

Categories were, initially introduced by Eilenberg and Mac Lane, around 1945 in a purely auxiliary fashion. Subsequently, the development of category theory was rapid, due to Lawvere's and numerous mathematicians' and logicians' work.

\subsection{Advantages of categories}

The idea about categories is that, since constructions with similar properties occur in completely different mathematical fields, we can capture those properties and drop the specific nature of each construction \cite{joy}. Categories require the least axioms and hence they are regarded as fundamental constructions. By its construction a category consists of objects and morphisms (arrows) between them; hence diagrams, showing the structure and the relations between objects.

A great advantage of categories is that we can visualize complicated facts by means of diagrams. Thus a categorical description helps to direct and organize one's thought (\cite{joy}).

 \subsection{Physical incentive}
 
 Nevertheless, the mathematical beauty and elegance of a theory, like category theory, does not suffices  to employ it, for constructing a physical theory. There are various reasons that, from a physical viewpoint, encourage us to modify the present framework of theories of physics. The motivations for doing so can be found in the original work by C. Isham, J. Butterfield and A. D\"oring \cite{AD & CJI I, roles, IB I}. 
 
 On one hand, there is the challenge of constructing a quantum theory of gravity. The discussion about the possibility of achieving a consistent quantum theory of gravity goes much beyond the scope of this paper. However in a topos framework for physics there seems to be enough space for such a theory. What is worth noting here, is the that an abandonment of continuous quantities, in a quantum gravity context, might be needed \cite{revolution}, and this is, actually, one of the features of topos physics, described here.

  On the other hand the conviction, that a restoration of a \emph{realist} model for quantum physics is possible, encourages one to go on with a topos description of physics. As we shall see later, we are able to construct a scheme for quantum physics, that looks much like the classical one.

\section{Ingredients}

Before going on a more precise description we give the `ingredients' - mathematical tools that we need in order to further our study. 
\subparagraph{Topos and category theory.} It is the main mathematical tool, that we employ here. The relevant parts of it are presented in the next chapter.
\subparagraph{Logic and algebras.} Logic is closely related to a topos. Classical and (specially) intuitionistic logic, and their representations, namely Boolean and Heyting algebras, are of great importance here.  The basics of logic and formal languages are presented in chapter three.
\subparagraph{Propositional calculus.} A formal language is closely related to a logic. Terms as `proposition' and `truth values' play a central role in the present paper. They are discussed in chapters three, four and five.
\subparagraph{Elements of functional analysis} such as operator spectra, von Neumann algebras and Gel'fand transforms are needed for the construction of `quantum topos'. Most of them are given in the appendix. \\

Last but not least a great amount of patience and a creative spirit is needed, in order to combine all those to a successful scheme for physics!
\newpage

\section{The general idea}
\subsection{Language of a system}
The present paper focuses on the original work of C. Isham and A. D\"oring \cite{AD & CJI I, AD & CJI II, AD & CJI III}. It is worth noting that this is the first time that topoi are employed to physics and the concepts proposed are completely novel. In those papers, a \emph{fundamentally new way of constructing theories of physics} is presented. As said before the main mathematical tool is topos theory. The idea is that to each physical system $S$ a formal language is associated \cite{AD & CJI I}. Two different languages are proposed: the propositional language $\PL S$ and the `local' language $\Ln S$. The general claim is that constructing a theory of physics is equivalent to finding a representation of that language in an appropriate topos. The topos employed by classical physics is the category of sets - just a special case of a category, where the objects are just sets and a morphism is a function between a pair of sets. Other type of theories employ different topoi.

This procedure might sound strange. But, once one gets familiar to the new mathematics, fascinating physics start coming out. The `universe' of topos is different than the `universe' of the ordinary sets. A topos carries its own world of mathematics; a world which, generally speaking, is not the same as that of classical mathematics \cite{AD & CJI I}. The (non Boolean) algebra that can `extracted' out  of a topos\footnote{more accurately: the sub-objects of an object in a topos form a Heyting algebra. All the necessary formal definitions are given in the following chapters} allows us to \emph{represent} a formal language to a topos.

\subsection{Propositions, states and quantity values}
In classical physics a scheme that works quite well is the following: the states of a physical system $S$ live in a state space $\S$, \ie a symplectic manifold. The physical quantities are real valued functions $A(s)$; they map a state $s$ of the system (a point in the phase space $\S$) to the real line $A: s \ra \mathR$.  Then a proposition about that system (for example ``the physical quantity $A$ of the system  has a value that lies in a subspace $\De$ of the real line $\mathR$'') holds \emph{true} if and only if the state $s$ of the system `makes' the physical quantity $A$ obtain that value. In a more clumsy way we say that if the state $s$ lies in the set $A^{-1}(\De)$, then the proposition $A \in \De$ is true, or $\nu(A \in \De)=1$. So a state $s$ assigns a \emph{truth value} to the proposition about $S$.

So everything works fine here; there is a well define phase space $\S$, a real line $\mathR$ and real-valued functions $A(s)$. The question now is: ``can we do a similar scheme for quantum physics?'' , $\ie$ define a state space  and real-valued functions from the states to the reals. 

Unfortunately Kochen-Specker theorem says a definite $no$, if the Hilbert space $\Hi$ has a dimension $dim\Hi \geqslant 3$. \ie there cannot be a generalized quantum mechanical phase space $\O$ such that an observable $A$ is given as a mapping $$f_A :  \O \ra \mathR$$ (named a \emph{hidden variable}) (A. D\"oring, \cite{andreas}).

Recall that in quantum mechanic the physical quantities are represented by self-adjoint operators in a Hilbert space $\Hi$, which $cannot$ regarded as a phase space. So Kochen-Specker theorem suggests the abandonment of a naive \emph{realistic representation} of quantum-physics. 

But, now let topos take over! In a topos formulation of quantum physics a realistic model is possible. 
As we shall see, in the context of the topos $\copv$ we can define 
\begin{itemize}
  \item a state-object $\Siu$; this is the analogue of a state space and can be naively described as a collection of many `local state spaces', 
  \item a truth-object $\underline{\mathbb{T}}$; the analogue of a state and 
  \item an arrow $\breve{\de}^o(\A)$ to represent a physical quantity, the analogue of $A$ (see fig. 1.1) 
\end{itemize}

Hence an analogy to the classical case is achieved\footnote{Those ideas are widely discussed in the main part of that work (chapters 5, 6 and 7).
}. Last in the place of $\mathR$ we now have the object $\underline{\mathR}^{\succeq}$; which now is a special function, named `presheaf', and not the real numbers for the values that physical quantities obtain! Welcome to the topos world!

\begin{figure}[h]
\begin{center}
\scalebox{0.8}{ \includegraphics{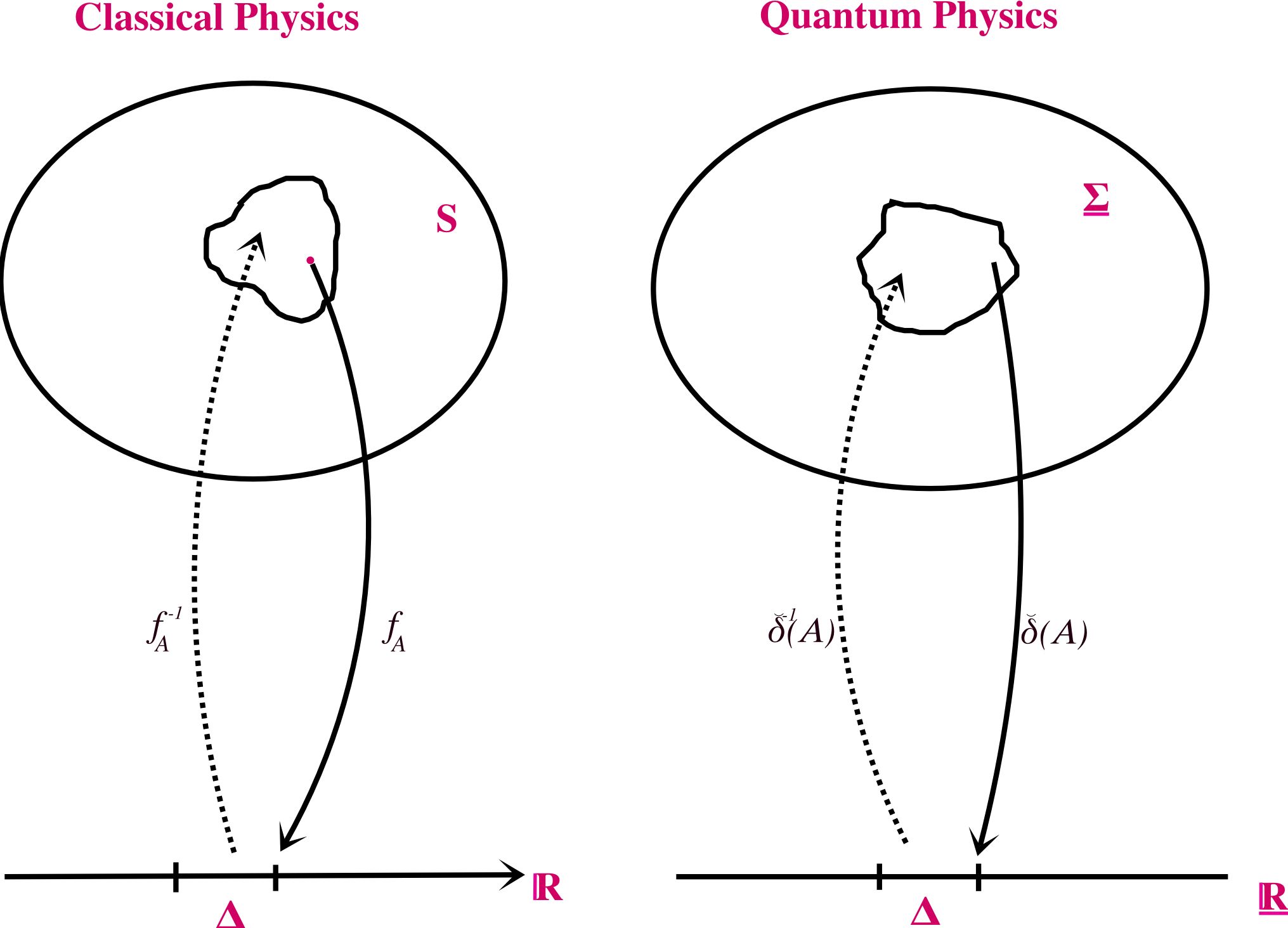} }
\caption{Propositions for classical and quantum physics}
\end{center}
\end{figure}

\newpage

Most of the material used in this paper has been taken from the recent work of C.J.Isham and A. D\"oring, \cite{AD & CJI I, AD & CJI II, AD & CJI III} and also the standard textbook for topoi, by Goldblatt \cite{Gold}, which I have interpreted and presented here, according to my own understanding of the subject.

\chapter{Categories and Topoi}   

\paragraph{A brief account of category theory}

\section{Introducing categories}
\emph{Category theory} is a quite general and abstract discipline of mathematics which deals with mathematical structures and relationships between them.  A \emph{category} can initially be conceived as universe of mathematic discourse; such a universe is determined by specifying a certain kind of objects and kind of ``functions'' between objects, that we call `arrows' or `morphisms'. As we shall see the notion of `duality' is present in every property of a category and plays a central role in its concept.

\subsection{Objects and arrows}

So a category, simply, consists of  
\begin{enumerate}
  \item a collection of objects 
  \item a collection of morphisms between these objects such that the fol lowing 
conditions hold:
\end{enumerate}

\begin{itemize}
  \item  composition condition: given two morphisms $f : a \ra b$ and $g : b \ra c$ 
with $dom~g=cod~f$ then there exists the composite map $g\circ f : a \ra c$  
  \item associative law: given a $a \mapright{f} b \mapright{g} c$ then $(h \circ (g \circ f)) = ((h \circ g) \circ f)$, such that the diagram:
  \begin{displaymath}
    \xymatrix{
        a \ar[r]^{f} \ar[d] \ar@{.>}[rd]^{h \circ g} & b \ar[d]^{g} \ar@{.>}[ld]_{g\circ f} \\
        d            & c \ar[l]^{h}}
\end{displaymath}
commutes
\item  identity law: for any object $b$ in the category there exists a morphism 
$1_b : b \ra b$ called \emph{identity arrow} such that, given any other two morphisms $f : a \ra b$ and $g: b \ra c$ we get $1_b \circ f = f$ and $ g \circ 1_b = g$, such that the diagram
\begin{displaymath}
    \xymatrix{
        a \ar[r]^{f} \ar[rd]_f   & b \ar[d]^{1_b} \ar[rd]^{g} &  \\
           &    b \ar[r]_g            & c }
\end{displaymath}
commutes. 

 \end{itemize}
 where `dom' and `codom' denotes the domain and codomain of a morphism, respectively.

\subsection{Examples of categories}
Category is a fundamental construction; indeed, it requires the least axioms. Hence it is a very wide and general mathematical construction, such that there is enough `space' for all many of the different mathematical structures.

Many mathematical theories can be adopted from category theory. If the features of a theory are the `input' in a category ($\ie$ objects and arrows), then the `output' will be the theory itself, or better, a categorial description of it.
For example group theory can be seen as a category, namely the category of groups \tb{Grp}, which has  groups as objects and the homomorphisms between those, as arrows. Similarly, a categorial descirption of linear algebra is captured in the category \tb{Vec}, which has objects the vector spaces and arrows between them, the linear transformations in that vector space. Similarly, we have the category \tb{Top} of topological spaces and continuous functions between them. Last, a category that plays special role in our work, is the category \tb{Sets}, of ordinary sets as objects and functions between sets as arrows.

\section{The category structure}
Since a category is defined there is a lot of algebra someone can do on it. Many properties, familiar from other disciplines of mathematics, are present in categories, in a more abstract way. We will give here only the most important, for what follows. A more thorough analysis can be found in the (classical) textbook for categories \cite{Gold}. Most of the definitions in that section comes form that book.
\subsection{Some definitions}
\subparagraph{A monic arrow $f: a \ra b$} between two objects, $a$ and $b$, in a category, is an arrow such that for any parallel pair $g, h: c \rightrightarrows a$ of arrows, the equality $f \circ g = f \circ h$ implies that $g=h$.  Usually, a monic arrow is denoted by $f: a \rightarrowtail b$.
\subparagraph{An epic arrow $f: a \ra b$} between two objects, $a$ and $b$, in a category, is an arrow such that for any parallel pair $g, h: b \rightrightarrows a$ of arrows, the equality $g \circ f = h \circ f$ implies that $g=h$.  Usually, an epic arrow is denoted by $f: a \twoheadrightarrow b$.

\subparagraph{An initial object} in a category is an object 0, if for every object $a$ in the category, there is one and only one arrow from 0 to $a$.
\subparagraph{A terminal object} in a category is an object 1 if for every object $a$ in the category, there is one and only one arrow from $a$ to 1. 
\subparagraph{An iso arrow $f: a \ra b$} is an arrow $f$ between two objects, $a$ and $b$, in a category, if there is  an arrow  $g: b \ra a$ such that $g \circ f = 1_a$ and $f \circ g = 1_b$.
\subparagraph{Isomorphic objects} denoted $a\cong b$ are said two objects, $a$ and $b$, if there is an arrow $f : a \ra b$ that is iso in the category, \ie $f: a \cong b$.
\subparagraph{A global element} of an object $A$, in a category, is defined to be an arrow $1 \ra A$. The collection of all global elements of $A$ is denoted $\Ga A$. A global element is not necessarily itself an object in the category.

 \subsection{The pullback}
 A \emph{pullback} of a pair $a \mapright{f} c \mapleft{g} b$ of arrows with a common codomain is a limit, in our category, for the diagram
 
\begin{displaymath}
    \xymatrix{
           & b \ar[d]^{g}   \\
           a \ar[r]_f &   }
\end{displaymath}

A \emph{cone} for this diagram consists of three arrows $f', g', h$ such that 

\begin{displaymath}
    \xymatrix{
            d \ar[rd]^{h}\ar[r]^{f'} \ar[d]_{g'}    & b \ar[d]^{g}   \\
           a \ar[r]_f &  c }
\end{displaymath}

 commutes. But this requires that $h = g \circ f' = f \circ g'$, so we may simply say that a \emph{cone} is a pair $$a \mapleft{g'} d \mapright{f'} b$$ of arrows, such that the ``square'' 
 
\begin{displaymath}
    \xymatrix{
            d \ar[r]^{f'} \ar[d]_{g'}    & b \ar[d]^{g}   \\
           a \ar[r]_f &   c}
\end{displaymath}
 commutes, \ie $f\circ g' = g \circ f'$.
 Thus, 
 \begin{definition}
a pullback of the pair $a \mapright{f} c \mapleft{g} b$ in a category, is a pair of arrows $a \mapleft{g'} d \mapright{f'}b$ such that 
\begin{enumerate}
  \item $f \circ g' = g \circ f'$, and
  \item whenever $a \mapleft{h} e \mapright{j} b$ are such that $f\circ h - g \circ j$, then
  
\begin{displaymath}
    \xymatrix{
        e \ar@/^/[rrd]^{j} \ar@/_/[rdd]^{h}  \ar@{.>}[rd]^{k} &                                     &                      \\
                                                            &     d \ar[d]^{g'}  \ar[r]^{f'}    &    b  \ar[d]^g    \\
                                                            &     a \ar[r]_{f}              &    c  
                                                             }
\end{displaymath}
  
  there is exactly one arrow $k: e \ra d$ such that $h=g' \circ k$ and $j=f' \circ k$. In other words when $h$ and $j$ are such that the outer ``square'' of the above diagram commutes, then there is only one way to fill in the arrow ($k$) to make the whole diagram commutes.  
\end{enumerate}
 \end{definition}

 \subsection{Sub-object}
\begin{definition}
A sub-object of an object $d$ of a category, is a monic arrow $f: a \rightarrowtail b$ with codomain $d$.
\end{definition}

The collection $Sub(d)$, of all sub-objects of an object $d$ is defined:  $$Sub(d) = \{[f]\}:~\mbox{ f is a monic arrow with $cod~ f$ = $d$}\}$$ where $[f]$ is the equivalence class $[f] = \{g: f\simeq g\}$.

 \subsection{Sub-object classifier}
\begin{definition}
If $\goth b$ is a category with a terminal object 1, the sub-object classifier for $\goth b$ is an object $\o$ of $\goth b$ together with an arrow, named $true$, $true: 1 \ra \O$ that satisfies the following axiom.

$\O-axiom$. For each monic $f: a \rightarrowtail b$ there is one and only one arrow in $\goth b$ such that 

\begin{displaymath}
    \xymatrix{
        a \ar[r]^{f} \ar[d]_{!} & d \ar[d]^{\chi_{\tiny f}} \\
        \bf{1} \ar[r]_{true}       & \Omega }
\end{displaymath}

is a pullback square
\end{definition}

The arrow `$!$' denotes a unique arrow in the category. The arrow $\chi_f$ is called the \emph{characteristic arrow}, or \emph{character} of the monic (\ie sub-object of $d$). The sub-object classifier, when it exists in $\goth b$, is unique, up to isomorphism (\ie there is an iso arrow between $\O$ and an object $b$).

\subsection{Power object}
A category $\mathcal{B}$ with products is said to have power objects if to each $\mathcal{B}$-object $
\alpha$ there are $\mathcal{B}$-objects $\cal P$($\alpha$) and $\epsilon_\alpha$, and a monic $
\epsilon : \epsilon_\alpha \rightarrowtail  {\cal P} (\alpha)\times\alpha$, such that for any $\mathcal{B}$-object b, and ``relation'', r: R $ \rightarrowtail b \times \alpha$ there is exactly one $\mathcal{B}$-arrow $
\emph{f}_r: b \rightarrowtail \cal{P}(\alpha)$ for which there is a pullback in $\mathcal{B}$ of the form

\begin{displaymath}
    \xymatrix{
        R \ar[r]^{r} \ar[d]_{} & b\times a \ar[d]^{f_{r} \times id _a}  \\
        \epsilon_a \ar[r]_{\epsilon}       & \cal{P}(\alpha)\times \alpha }
\end{displaymath}
\\

\section{Functors and Presheaves}
\subsubsection{Functions between categories}

We have defined categories as collections of objects and arrows between them. This construction is 
quite general and abstract so there is `space' for many applications of it, as said before. A natural 
question that comes up out of categories is if (and how) one can move from one category to the other. 
Keeping in mind that
in category theory it is the morphisms, rather than the objects, that have the primary 
role we can take a more global viewpoint and consider categories themselves as structured objects. 
Now, if we define an arrow between those objects we are done! It turns out that we can construct such 
morphisms, but with some restrictions, regarding the arrows of the starting and the resulting category.
The ``morphisms" between them, that preserve their structure, are called \emph{functors}. To be more 
precise we should move to the formal definition of a functor:

\begin{definition}
A \tb{covariant functor} $\tb F$ from a category $\cA$ to a category $\cB$ is a function that assigns to 
each $\cA$-object A a $\cB$-object ${\tb F}_A$, and to each $\cA$-morphism  A $\mapright{f} A^{\prime}
$ a $\cB$-morphism  
${\tb F}_A \mapright{{\tb F}(f)} {\tb F}_{A^{\prime}}$ in such a way that 
\begin{enumerate}
  \item $\tb F$ preserves composition; \ie ~$${\tb F}({f\circ g})={\tb F}({f})\circ{\tb F}({g}),$$ for two 
morphisms $g:\tilde{A}\rightarrow A$, $f: A\rightarrow A^{\prime}$,  whenever f$\circ$g is defined, and
  \item $\tb F$ preserves identity morphisms; \ie ~${\tb F}_{id_A}=id_{{\tb F}_{A}}$ for each $\cA$-object 
A.
\end{enumerate}
 
\end{definition}
This definition comes from Ad$\grave{a}$mek, Herrlich and Strecker \cite{joy}, but is more or less the same in all textbooks.

\begin{definition}
A \tb{contravariant functor} $\tb X$ from a category $\cA$ to a category $\cB$ is a function that assigns to 
each $\cA$-object A a $\cB$-object ${\tb X}_A$, and to each $\cA$-morphism A $\mapright{f} A^{\prime}
$ a $\cB$-morphism  
${\tb X}_{A^{\prime}} \mapright{{\tb X}(f)} {\tb X}_{A}$ in such a way that 
\begin{enumerate}
  \item $\tb X$ inverses composition; \ie ~$${\tb X}(f\circ g)={\tb X}(g)\circ{\tb X}(f),$$ for two morphisms 
$g:\tilde{A}\rightarrow A$, $f: A\rightarrow A^{\prime}$,  whenever f$\circ$g is defined, and
  \item $\tb X$ preserves identity morphisms; \ie ~${\tb X}_{id_A}=id_{{\tb X}_{A}}$ for each $\cA$-object 
A.
\end{enumerate}
 
\end{definition}

\begin{definition}
A \tb{presheaf} $\underline{X}$ on a category $\cal C$ is a covariant functor $\underline{X} : {\cal C}^
{op} \rightarrow \tb{Sets}$. Alternatively, it is a contravariant functor $\underline{X} : {\cal C} \rightarrow 
\tb{Sets}$.

\end{definition}

Roughly speaking a presheaf is a set which varies from one context to another.
A definition and a discussion of a presheaf in a poset, can be found at the appendix. 
\newpage
\begin{displaymath}
    \xymatrix{
        A \ar[r]^{F} \ar[d]_f & F_A \ar[d]^{F(f)} \\
        A' \ar[r]_{F}       & F_{A'} }
\end{displaymath}
\begin{center}
$\cal{C} \mapright{F} \cal{D}$
\end{center}

{\small \textit{A Functor maps a category $\C$ to another category $\cal{D}$, preserving the morphism 
order.}}

\begin{center}
\begin{displaymath}
    \xymatrix{
        A \ar[r]^{\X}                  & \X_A \ar[d]^{\X(f)} \\
        A' \ar[r]_{\X}  \ar[u]_f   & \X_{A'} }
\end{displaymath}

${\cal{C}}^{op} \mapright{\X} \textbf{Sets}$

\end{center}

{\small \ti{A Presheaf maps a category $\C$ to the category \tb{Sets} of sets}} \\

A more `complete' diagram, showing a covariant functor from a category $\C$, with objects \{$A,B,C,D 
\ldots$\} and arrows \{$f, g, h, k \ldots$\} to a category $\cal D$ with objects \{$F_A. F_B, F_C, F_D \ldots$
\} and arrows \{$F(f), F(g), F(h), F(k) \ldots$\} is the next one 

\begin{displaymath}
    \xymatrix{
                                     &      &  B \ar[lld]_h  \ar@{~)}[d] \ar[r]^g & C \ar[lld]      \ar@{~)}[d] \\
           A \ar@{~)}[d]^{F} \ar[r]_f    &   D  \ar@{~)}[d]^F         & F_{B} \ar[r]^{F(g)}  \ar[lld]&   F_C \ar[lld]^{F(k)}          
\\
           F_A \ar[r]_{F(f)}             &   F_D 
        }
\end{displaymath}

where the upper square is the category $\C$, the lower one is the category $\cal D$ and the curly arrows 
depict the covariant functor $F$.

\section{Definition of topos}

Roughly speaking a topos is category that behaves much like \tb{Sets} \cite{AD & CJI I} or, sometimes, it is modeled after the properties of the \tb{Sets}. Its notion captures some of the aspects that are defined in sets, such as the union and join (products and co-products).

 Formally it is defined as follows.

\begin{definition}
An \emph{elementary topos} is a category $\cal E$ such that 
\begin{enumerate}
  \item  $\cal E$  is finitely complete,
  \item  $\cal E$  is finitely co-complete,
  \item $\cal E$ has exponentiation, 
  \item  $\cal E$  has a sub-object classifier. 
\end{enumerate}
\end{definition}

An alternative definition can be given, using power objects.

\emph{A category $\cal E$ is a topos if and only if  $\cal E$ is finitely complete and has power objects}. (\cite{Gold}, p. 106)

Unfortunately the lack of space does not allow us to present what of (co-)completeness and exponentiation are. The reader is prompted to \cite{Gold}, p. 69-71.

\chapter{Logic, Algebras and Topoi}    

Giving a brief explanation of what a logic is, \ti{is not an easy and perhaps not useful matter}\footnote{quote by E.J. Lemmon, author of numerous books in logic.}. Generally speaking, a logic is any set of rules for forming new sentences from given ones (the logic's axiomatics) together with rules for assigning truth values to them (the logic's semantics). Roughly, a logic is a set of abstract `ideas', which can be realized (more concretely) when represented in an algebra.
We will shortly describe the two main types of logic - classical and intuitionistic - , give representations of them and show the connection to topos theory.

\section{Classical Logic}
Classical logic is computationally the simplest of all the major logics. It is , maybe, the most usual and well-studied one; hence its name.The semantics of classical propositional logic can be described just in terms of simple operations of 0's and 1's (usually captured by tables). Formally, a \emph{system} for classical logic, can be defined as an \emph{axiom system} whose axioms are given in the first part of appendix. But be before that, we have to go on with the appropriate description of the logical operations. Most of what follows, in that section, comes from \cite{Gold}, ch. 6.

\subsection{Simple and compound propositions}

A \emph{proposition} or (\emph{sentence} or \emph{statement}) is simply an expression that is either true or false. Thus:
$$``1+2=3"$$  $$``34~minus~3~equals~4"$$ $$``London~is~in~France"$$ are examples of propositions, while $$``Is~3+0~the~square~root~of~9?"$$ and $$``Do~not~ read ~that~ paper!"$$ do not count as propositions.

So a proposition is an entity that can be assigned one of the two \emph{truth-values} 0 and 1, \ie the truth values lie in the set $\tb{2}=\{0,1\}$. We assign the value 1 to a proposition if it is \emph{true} and the value 0 if it is \emph{false}. 

The logical operations $``and",~``or",~``not" $ serve as \emph{logical connectives} in constructing compound sentences, out of simple ones. Let $a,b,c$ be sentences. Then $``a~and~b"$, $``a~or~b"$, $``\sim a"$ are also sentences. We use the symbols $``a\wedge b"$, $``a\vee b"$, $``\sim a"$ for these operations, which are said to be obtained by $conjuction$, $disjunction$ and $negation$, respectively.

\subsection{Truth values of propositions}
We can assign truth values to compound propositions, made using the connectives $\wedge,\vee, \sim$, from the truth values of its components.

\begin{itemize}
\item \emph{Negation}  $\sim$ \\ 
The sentence $\sim a$ is true (assigned 1) when the sentence $a$ is false and false (0) when $a$ is true. Alternatively we can regard it as determining a function:
$\neg : \bf{2} \rightarrow \bf{2}$ defined by $\neg 1 = 0,~~ \neg 0 =1$. It is called the \emph{negation truth-function}.

\item \emph{Conjunction} $\bigcap$ \\
The truth-value of the compound proposition $a\wedge b$ is true if and only if both $a$ and $b$ are true. The $conjuction~truth-function$, from pairs of truth-values to truth-values:
$\cap: \bf{2}\times 2 \rightarrow 2$
is define by $1 \cap 1 =1,0 \cap 1 =1 \cap 0 = 0 \cap 0 =0$  

\item \emph{Disjunction} $\bigcup$ \\
A compound proposition $a\vee b$ is true if and only if either the propositions $a$ or $b$ or both are true. The $disjuction~truth-function$, from pairs of truth-values to truth-values:
$\cup : \bf{2} \times 2 \rightarrow 2$
is defined by  $1 \cap 1 =0 \cap 1 =1 \cap 0 = 1, 0 \cap 0 =0$  

\item Implication $\Rightarrow$\\
The implication connective allows us to form a sentence $``a~ implies~ b"$, symbolised $``a \supset b"$.
The proposition $``a\supset b"$ is false if $a$ is true while $b$ is false (we cannot derive something false from something true!). In all other cases $a\supset b$ is true.
The $implication~truth-function$ $\Rightarrow : \tb{2}\times \tb{2} \ra \tb{2}$ has $1\Rightarrow 0 = 0, 1\Rightarrow 1 =  0\Rightarrow 1 = 0\Rightarrow 0 = 1$

\end{itemize}

We the logic operations and truth functions we can assign truth-values to any compound proposition, just by applying successively the rules given above. There numerous properties that classical logic has. A thorough presentation of them is beyond our scope. The reader is prompted to \cite{Gold, sieves, sep}.

\section{Non-classical logic}
Broadly speaking the classical logic is the `classical' way we apprehend the world in our daily lives. It is natural to think that the proposition $$either~It~will~rain~tomorrow~or~it~will~not$$ is a tautology, \ie something always valid, in the context of the language used (here classical).   
However there are several systems of logic where the above expression does not hold true. 

With the term \emph{non-classical logic} we refer to a wide family of logics that differ from the classical one, in various aspects. The most usual non-classical logic are characterized by the lack of one or more of the following properties: \emph{the law of the excluded middle} $a \vee \neg a =1$, \emph{the law of non-contradiction} $a \wedge \neg a =0$, \emph{the commutativity of conjunction} $a \wedge b = b \wedge a$ and others, ($a, b$ are propositions).

As we shall see, a particular non-classical logic, the \emph{intuitionistic} or \emph{constructivist logic}, plays an important role in the present work. 

\section {Formal Language \label{PL}}
A Formal Language, \tb{PL} is normally given by an alphabet together with some formation rules\footnote{Sometimes a language over an alphabet is defined as set of strings made from the symbols in the alphabet. A string of length \textit{n} on an alphabet \textit{l} of \textit{m} characters is an arrangement  (grouping) of \textit{n} not necessarily distinct symbols from \textit{l}. There are \textit{m$^n$} such distinct strings. For example, the strings of length \textit{n}=3 on the alphabet \textit{l}=\{1,2\} of two characters are \{1,1,1\}, \{1,1,2\},\{1,2,1\}, \{1,2,2\}, \{2,1,1\}, \{2,1,2\}, \{2,2,1\}, and \{2,2,2\}.}.
 The idea of a formal language is directly connected to that of a logic; a language follows the rules of a particular logic, and allows us to extend the study of the calculus of propositions and truth values.

\paragraph{Ingredients of PL}
\subparagraph{Symbols}
\begin{enumerate}
\item infinite list of symbols $\pi_0, \pi_1, \pi_2 \ldots $, called propositional variables (or sentence 
letters)
\item symbols $\sim, \wedge, \vee, \supset$
\item bracket symbols ),(
\end{enumerate}

As we shall in the following chapter, a symbol of the language $\PL S$, for a physical system $S$, is simply the proposition $``A~\ve~\De" := \pi$, about that system. 
\subparagraph{Formation rules for \tb{PL} sentences}
\begin{enumerate}
  \item Each sentence letter $\pi_{i}$ is a sentence
  \item If a is a sentence so is $\sim$a
  \item If $\alpha$ and $\beta$ are sentences then so are ($\alpha\wedge\beta), (\alpha\vee\beta), 
(\alpha\supset\beta)$  
\end{enumerate}
Here the letters $\alpha, \beta$ represent generic sentences, which well could be a single (language-) 
letter or something more complicated like $((\pi_{1}\vee\pi_2)\supset(\pi_2\wedge\sim\pi_2))$. The 
collection of sentence of letters is denoted $\Phi_0$ while $\Phi$ stands for the set of all sentences:
 $$\Phi_0 = \{\pi_0, \pi_1, \pi_2 \ldots\} $$
$$\Phi = \{\alpha : \alpha~ is~ a~ PL-sentence\}$$
To develop a theory of meaning or \emph{semantics}, for PL we use the truth-functions V. 

By \emph{value-assignment} V:$\Phi_0 \rightarrow \bf{2}$ we mean any function from $\Phi$ to \{0,1\}$
\equiv$\tb{2}. \tb{Such a V : $\Phi_0 \rightarrow \bf{2}$ assigns a truth value V($\pi_i$) to each sentence 
letter and so provides an \emph{``interpretation''} to the members of $\Phi_0$.
} \\
\\
The interpretation can be systematically extended to all sentences, so that V extends to a function from $
\Phi$ to \tb{2}. This is done by induction over the formation rules through successive applications of the rules:
\begin{enumerate}
  \item $V(\sim a)=\neg V(a)$
  \item $V(a \wedge b) = V(a) \cap V(b)$
  \item $V(a \vee b) = V(a) \cup V(b)$
  \item $V(a\supset b)=V(a)\Rightarrow V(b)$
\end{enumerate}

In this way any $V: \Phi_0 \ra \tb 2$ is lifted in a unique way to become a function $V: \Phi \ra \tb 2$.

A sentence $a \in \Phi$ is said to be a $tautology$ if holds true for every assignment, \ie~ for every value assignment $V$, $V(a)=1$.

\section{Axiom systems}
\subsection{Axiom system for classical logic}
Apart from semantics, axiomatics is a needed part in constructing a language PL. These are concerned with deriving new sentences from given ones. More precisely we define an axiom system as:
\begin{enumerate}
  \item a collection of sentences, called \emph{axioms} of the system
  \item a collection of\emph{ rules of inference} which prescribe operations to be performed on sentences, to derive new ones. 
\end{enumerate}

A \emph{theorem} is a sentence that is derivable from the axioms.
The classically valid sentences are the ones whose theorems are precisely the tautologies of PL. The classically valid sentences are not obtained by a unique axiom system; indeed we can use several different axiom systems to deduce  the theorems of PL.

An axiom system, commonly used is called CL. All of its axioms can be derived from twelve forms that are listed in the appendix. The last axiom 
\begin{equation}
\label{excluded}
a \vee \neg a = 1
\end{equation} 
will be of particular interest.

It contains only one rule, the \\
\tb{rule of detachment:} \emph{from the sentences $a$ and $a \supset b$ the sentence $b$ can be derived.}
 (sometimes called `modus ponens').

\subsection{Axiom system for intuitionistic logic}

We introduced non-classical logic before. Now we are ready to give a more concrete description of the intuitionistic logic, by defining an axiom system, called IL, for it. The system is based on the same language PL. Its sole rule is also the rule of detachment and its axioms are almost of the same form as of the classical logic; there is one single form of CL that is not an axiom in IL. This is the axiom \ref{excluded}, which is called the \emph{law of the excluded middle}.

However, there are more CL tautologies that are not IL-theorems. The forms $$a \vee \neg a~~ \mbox{and} ~~\neg \neg a \Rightarrow a,$$ are \tb{not} theorems of IL, while $$a \Rightarrow \neg \neg a,$$ \tb{is} a IL-theorem.

  Intuitionistic logic is a part of classical logic, in the sense that all formulas provable in intuitionistic logic are also provable in classical logic. Although, some basic theorems of classical logic do not hold in intuitionistic logic.


To summarize, we write:
\subparagraph{ ~classical logic : $\quad a \Leftrightarrow \neg \neg a$ \\
intuitionistic logic :~~~ $a \Rightarrow \neg \neg a$} 

\section{Representation of a language}{\label{repr}}
Logic and languages, studied so far, are quite abstract constructions. But not the only ones! In various disciplines of mathematics, that study abstract constructions, there are propositions, the so-called \textit{representation theorems}, that establish an equivalence between the model of a certain abstract structure and a particular list of concrete models (\cite{Gold}, p. 26).

Following that idea, we can find such `concrete' models to represent on them a particular logic; in our case classical or intuitionistic.

\paragraph{A Boolean algebra} is a canonical structure in which we can represent a formal language, following the rules of classical logic.
\paragraph{A Heyting algebra} is canonical structure in which we can represent a formal language, following the rules of intuitionistic logic.\\

The definitions of a Heyting and Boolean algebras are a bit more technical (all the necessary definitions can be found in the appendix, \ref{lattices}). They are both defined as distributive lattices. The difference lies in the complement; a Boolean lattice is always \emph{complemented}, while the lattice of a Heyting algebra is \emph{pseudocomplemented}. This reflects the non validity of the statement $a \vee \neg a$ \ref{excluded}.

\subsection{A representation of Boolean algebra}
Since 1936 and due to ``Stone's representation theorem for Boolean algebras'', it is known that a connection between Boolean algebra and the subsets of a set can be established; the operations of classical logic can easily be represented by set-theoretical operations. If $A, B$ are two sets, let the operation $A\wedge B$ be represented by the meet $A \cap B$ of the two sets, $A \vee B$ by the join $A \cup B$, $\neg A$ by the complement $A^C$ and the implication $A \Rightarrow B$ by the set inclusion $A \subset B$. And that's it! The set-theoretical operations reproduce exactly the axioms of the classical logic. For example, the Boolean-logical assertion that a statement $a$ and its negation $\neg a$ cannot both be true,
   $$ a\land(\lnot a) = 0,$$
parallels the set-theoretical assertion that a subset A and its complement $A^C$ have empty intersection,
   $$ A\cap(A^C) = \varnothing.$$
   So, finding a representation to a Boolean algebra is equivalent to finding a representation to the subsets of set\footnote{Here, by sets, we refer to genuine sets and not sets of more complicated entities, as sieves or functionals, as we shall see later.}. Later on we will see the connection to classical physics.

For example the set {\textbf{2}}, which is the set of the truth values, in a classical logic, together with truth functions $\neg, \cap, \cup$ forms a \emph{Boolean Algebra}.

\subsection{Boolean vs Heyting algebra} 
 
 A similar representation of a Heyting algebra is not straightforward and easy to visualize. However we can give an example, pointing out the difference between operations in Boolean and Heyting algebras.

Let \emph{X} be a measurable space\footnote{A set considered together 
with the $\sigma$-algebra on the set}  and  \emph{a}$ \subseteq$ \emph{X} a  subset  of $X$.
In the case of Boolean algebra the logical negation $\neg\emph{a}$ of the proposition \emph{a} is represented by  the (ordinary set-theoretical) complement \emph{X}$\slash$\emph{a}.

Now let $X$ be a topological space\footnote{A set X together with a collection of open subsets T.}. 
 In the case of a representation of a \emph{Heyting algebra} the set \emph{a}$ \subseteq$ \emph{X} must be \emph{open}. The logical negation $\neg\emph{a}$  is defined to be the \emph{interior} of the complement \emph{X}$\slash$\emph{a}. Therefore the difference between $\neg\emph{a}$ in the the two cases is just the `thin' boundary of \emph{X}$\slash$\emph{a}.

\section{Algebra of the sub-objects and representation in a Topos $\tau$}
\sectionmark{Algebra of the sub-objects}

Now how all these are connected to a topos? In fact topos theory, since its foundations, served as novel and successful mathematical tool of studying logic. Every topos carries its own logic; an `internal' logic. Additionally it can be shown that the sub-objects of an object in a topos form an algebra.
And since, in the present work, we are looking for representations of a formal language in an algebra, we are justified to represent a language to a topos. A deep analysis of the algebra that sub-objects of an object in a topos, form cannot be given here (there is a whole chapter devoted to that, see \cite{Gold}, ch. 7). However a general result that is of great interest is that
in any topos the collection of the sub-objects of \emph{any} object $d$, together with an inclusion 
relation forms a poset (a partially ordered set). If this poset is a Boolean Algebra then our topos 
is called Boolean. However, in the general case \\
\emph{\tb{The set of all sub-objects of an object in a topos is a Heyting Algebra.}}\\

As said before, we can associate the rules of classical logic with the operations on subsets of a set. This idea can be extended in any topos; but it turns out that the logic to be represented is not classical anymore. The algebra of the sub-objects of an object, in a general topos, is a Heyting algebra and the logic represented is the intuitionistic.

\subsubsection{Truth function as arrows}

 To represent a language in a topos we must use the sub-objects of an object in that topos as the `letters' of our language. Then what's left is to define, are the `formation rules'. To do so we use the only ``interaction''  between (sub-)objects that we have; Arrows! Indeed, in the classical case, each of the truth-functions has codomain \tb 2 and so is the characteristic function\footnote{The characteristic function for a subset $ A$ of a set $ X$ is the function\\
$\displaystyle \chi_A(x) = \begin{cases}1,&\text{if }x\in A,\\ 0,&\text{otherwise } \end{cases}$} of some subset of its domain. Since the characteristic functions are special arrows in a topos we can succeed an arrows-only definition of the truth functions.
 In other words we are looking for an association of the logic connectives $\wedge, \vee, \sim, \supset$ to the arrows of our topos. 
 
These can be found in an explicit form in \cite{Gold}, p. 136-140.

In the topos of sets it is easy to see how the logic operations are defined on characteristic functions. The following theorem establishes a relation between set operations and truth functions.

\begin{theorem}
If A and B are subsets of D, with characteristic functions $\chi_A : D \rightarrow \mbox{\tb{2}}$, $\chi_B : D \rightarrow$ \tb{2} 
then
\begin{enumerate}
  \item $\chi_{\neg A}$ = $\neg\circ\chi_A$
  \item $\chi_{A\cap B} = \chi_A \cap \chi_B$
  \item $\chi_{A\cup B} = \chi_A \cup \chi_B$
\end{enumerate} 

\end{theorem}

The proof of that is quite easy and can be found in Goldblatt, \cite{Gold}, p. 146. Furthermore, one can defines truth arrows in a topos and go on with the semantics on it, define operations (intersection, union, complement) on the collection of sub-objects $Sub(d)$ of an object $d$. However we will not go that far. We just keep the result that ``the set of all sub-objects of an object in a topos is a Heyting Algebra''  and so a topos can be used for representing a formal language on it, for our further study.

After this brief introduction we are ready to see how can topos theory serve as tool in constructing physical theories.

\chapter{Topos theory as a framework for theories of physics \label{ch5}}     
\chaptermark{Topos theory as framework}

\section{Introduction}
So far we have introduced some abstract constructions; categories. topoi and the related algebras on 
them. But how can these be used to describe physical systems?

As mentioned earlier, the idea is that using topoi we can achieve a generalization of the ordinary set 
theory that lies underneath today's physical theories. Roughly speaking we substitute the objects `sets' 
in our topos with objects other than `sets'. Then we are looking for a physical representation of a formal language on it. But before going further, let's stick for a while in the logic of the classical physics; see how a familiar description of a classical system is seen from a linguistic aspect and how the whole scheme can be nicely captured by a topos.

\paragraph{Property of the System.}
The reason why we need a physical theory is that it specifies how a physical quantity (\ie observable) 
acquires a value. The scheme that can be used is that the physical quantity is actually a `mapping' from 
the \emph{state space} (where the state of system lives) to the \emph{value space}.

\begin{displaymath}
    \xymatrix{ 
{\begin{tabular}{|l|l|}
  \hline
  {\small state space} \\ \hline
\end{tabular}} 
~ \ar[r]^{physical}_{quantity} ~~~~~& ~~~~~~~~~ {\begin{tabular}{|l|l|}
  \hline
  {\small quantity-value space} \\ \hline
\end{tabular}} }
\end{displaymath}

 The value of the physical quantity is a `property of the system' and hence must be meaningful and 
representable in the theory. In our case the physical quantities are represented by arrows whose 
domain is the state space $\S$ and codomain the value space. In the classical case $\S$ is a set, but this is not true for theories like quantum physics. 

All these can be realized in the topos language. 
\begin{itemize}
\item The state space and quantity-value space are specific objects of the topos, called `state object'  $
\Sigma$ (topos analogue of the classical state-space) and `quantity-value object' $\mathcal{R}$, respectively
\item Physical quantities are represented by morphisms in the topos
\item Propositions\footnote{also called `statement', in the context of a formal language} are represented by sub-objects of the state object $\Sigma$
\end{itemize}

\begin{displaymath}
    \xymatrix{ 
{\begin{tabular}{|l|l|}
  \hline
  {\small $\Sigma$} \\ \hline
\end{tabular}} 
~ \ar[r]^{physical}_{quantity} ~~& ~~~~~~~~~ {\begin{tabular}{|l|l|}
  \hline
  {\small $\mcl{R}$} \\ \hline
\end{tabular}} }
\end{displaymath}
\\

Now in the case of Classical Physics the topos is just the category of sets, \textbf{Sets} where:
\begin{itemize}
\item the state space $\mathcal{S}$ is a set
\item physical quantities are represented by real-valued functions $\breve{A}: \mathcal{S} \ra \mathR $
\item propositions are represented by subsets of $\mcl{S}$ 
\end{itemize}

Now, since we can associate the propositions about a system with the subsets (or sub-objects of an object in a topos, in general)  we expect them to form an algebra. We have already pointed that the subsets of sets, form a \emph{Boolean algebra} (alternatively, the topos \tb{Sets} is Boolean). However for a generic topos, as we shall see, this is not the case.
We see that from our general construction one can easily deduce the classical system and logic, just by 
using \textbf{Sets} as the topos. The topos representation of physics seems to be a promising way of 
translating physical theories! What is of importance here is that the value object $\mcl{R}_{\phi}$ need 
not be the set of real or even complex numbers. Since there is no \emph{prima facie}\footnote{based 
on the first impression; accepted as correct until proved otherwise} reason for requiring the value space 
to be continuum, we will not restrict ourselves to the case $\mathcal{R}_{\phi} \equiv \mathR$. As we 
shall see this will be a fruitful assumption, or better, a fruitful rejection of a (previous) assumption!

\section{Constructing Physics in a topos}

Our general task is to represent a physical system (or some features of it) by structures in a suitable topos. The claim is that for a given physical system, described by a specific theory-type (for example classical physics or quantum physics) there is a specific topos $\tau_{\phi}(S)$ related, where $S$ is associated with the system and the subscript $\phi$ with the theory-type. That topos construction can captures the features of the physical system $S$ and, as we shall see,  disclose `new' ones.

 We have seen so far that each topos carries its own logic and this allows us to link it with a formal language. 
More accurately if we find a language, appropriate for a system described by a theory-type, we can represent it in a topos. Deriving a physical theory is equivalent to finding a representation of a language in a topos. As we shall later see for the theory-type `quantum physics' we find an appropriate language, represent it in a (Heyting) algebra and hence to the sub-objects of a (specific) object in the topos, that obey this algebra.
 
 \subsection{The Propositional Language $\PL S$}
 To start dealing with languages in topoi, let's take the collection of \emph{propositions}. The resulting 
set $\PL S_0$ of all strings of the form ``A$\varepsilon \Delta$'', where A is some physical quantity of the system \emph{S} and $\Delta
$ is a (Borel) subset of the real line $\mathR$. Now this set can be `promoted' to a language if equipped 
with some rules. To do so we add a new set of symbols ($\neg, \wedge, \vee, \Rightarrow$) and then we 
define a sentence inductively, in accordance to paragraph \ref{PL}:
 \begin{enumerate}
  \item Each primitive proposition ``A $\varepsilon~\Delta$'' in $\PL S_0$ is a sentence 
  \item If $\alpha$ is a sentence then so is $\neg\alpha$ 
  \item If $\alpha$ and $\beta$ are sentences then so are $\alpha\wedge\beta, \alpha\vee\beta$ and $
\alpha \Rightarrow \beta$ 
\end{enumerate}

The collection of all sentences in $\PL S$ forms an elementary formal language. Notice that in this 
language the quantifiers `$\forall$' and `$\exists$' are not contained. Hence we say that $\PL S$ is a 
\emph{propositional} language only.

\subsection{Representations of the language}
The rules defined above are not concrete at all; they have no explicit meaning. If we want the abstract 
symbols $\neg, \wedge, \vee, \Rightarrow$ yield an actual meaning  we have to map them in a less abstract structure. Following the ideas of \ref{repr}  we proceed with a \emph{representation} of the language in an algebra. In the most general case this algebra is a Heyting algebra $\mathfrak{H}$.

\tb{\textit{A representation $\pi$ maps each of the primitive propositions, $\alpha$ in $\PL S_0$ to an 
element $\pi(\alpha)$ of some Heyting algebra.
$$\pi:~ \PL S \rightarrow \goth{H}$$}}

Since different systems can have the same language, some of the features of a theory might lie in the 
language (linguistic precursors) while others might exist only in the representation of it.  The question, when representing a language to construct a theory of physics, is how much information 
we should encompass in the language. We shall return to that later. 

From general topos theory we know that the sub-objects of an object in a topos is a Heyting algebra (or Boolean as a special case). This leads us to look for a representation of a language in a topos. In the context of the chosen language what matters is the ability of writing propositions that obtain a truth value. This can be done when `states' are defined, within the representation of the language. States yield a truth value for the primitive propositions in $\PL S$. In theory-type `classical physics' the topos is always \tb{Sets} for any physical system. In such a case the algebra that sub-objects ($\ie$ subsets) of objects ($\ie$ sets) of \tb{Sets} form, is the ordinary Boolean algebra. This means that the propositions in the language, that is to be represented in \tb{Sets}, will either be \emph{true} or \emph{false}. As we will see this is not happening in a quantum topos, which of course is not \tb{Sets} and the algebra is not Boolean.

\paragraph{A `concrete' proposition} 
To make these clearer we will give a (baby) example. We regard as our system $S$ a point-particle moving in one dimension. The Hamiltonian describing $S$ is  
\begin{equation}
\label{example}
H = \frac{p^2}{2m} + V(x)
\end{equation}

The state space is the phase space $\mathR^{2n}=\mathR^n\times\mathR^n$, which is a symplectic 
manifold\footnote{a symplectic manifold is a pair (M, $\omega$), where M is a manifold and $\omega$ is 
a symplectic form} $(M, \omega)$. In the one-dimensional case, of course, we have $\mathR^{2n}\equiv
\mathR^2$. The position $q$ and momentum $p$ of the particle, lie in $(M, \omega)$.

A physical quantity, \emph{energy} for example, is represented by the real-valued function $H : \cal{S} 
\rightarrow \mathR$. Now what can we say about that system? Make a \emph{proposition} about it! For example the form
 $$H(p,q)=\vr $$ 
means (\emph{proposes}) that the energy of $S$ has a value in $\vr$. 
 where $p, q \in \cal{S}$ and $\varrho \in \De$. So, and in 
accordance to our prior knowledge of the classical Hamiltonian systems, \emph{H} maps the pair $(p,q)$ to a real number, which is the value of the energy. Hence the physical quantity $energy$  is represented by a real-valued function $H$.

 A proposition $``H~\ve~\De"$ about the system is represented by the subset of states of the phase 
space for which the physical quantity lies in  the subset $\Delta$ of the reals;
 $$ \pi_{cl}(H~\ve~\De) = H^{-1}(\De)~\subseteq \mathR^2$$ (where we droped the hats `` $\breve{}$ '' )

Since the representation takes place in \tb{Sets} the logic of the propositional calculus in theory-type 
`Classical Physics'  is represented in a Boolean Algebra.

After that, we are ready to see in detail how different physical theories are connected to a language 
representable in a topos. 

\begin{chapter}
{Representing in a Topos}   
\label{chap}
\end{chapter}

\section{From representations to physical theories}

As mentioned before, our general claim is that a physical theory is equivalent to finding a representation 
of a typed formal language, that is attached to the system, to an appropriate topos.  

But before going further to see how classical and quantum physics arise from topoi, lets give some 
examples in order to emphasize the underlying connection between sets, topoi, languages and logic.

In chapter 4 we dealt with propositions, about a system in classical physics, of the form `A$\varepsilon\Delta$', which state that the 
value of the physical quantity A lies in the subset $\Delta$ of the real line $\mathR$. The propositions 
are represented by subsets $\breve{A}^{-1}(\Delta) \subseteq \mathcal{S}$, where $\breve{A}$ is a real-
valued function\footnote{To be more precise, we are interested in \emph{measurable functions} and \emph{Borel subsets}} associated with the physical quantity A.  Actually any proposition P about the system is represented by an associated subset $\mathcal{S}_P$ of $\mathcal{S}$ and conversely every subset of $\mathcal S$ represents a proposition P. We will, shortly, see the analogue in the case of quantum physics.

The logical calculus arises naturally when someone considers the proposition ``P and Q" or ``P or Q''. 

Propositions P and Q are now represented by the subsets $\mathcal S_P$ and $\mathcal S_Q$ 
respectively. Let's  form the `composite' proposition ``P and Q''. We note that this proposition is true if and 
only if both P and Q are true. This means that the states that are represented by P and Q lie in both sets, 
i.e. in the set-theoretical intersection of $\mathcal{S_P}$ and $\mathcal{S_Q}$. Hence $\mathcal S$$_
{(P~ and~ Q)}$ = $\mathcal{S_P\cap S_Q}$ and 

$$``P~and~ Q"~ is~ represented~ by~ \mathcal{S_P\cap S_Q}$$ 

Similarly we find that 

$$``P~or~ Q"~ is~ represented~ by~ \mathcal{S_P\cup S_Q}$$ 

and $$``not~P"~ is~ represented~ by~ \mathcal{S/ S_P}$$ 

The composite proposition ``P or Q'' is true if either or both of P, Q are true, i.e. all the states that lie in 
the set-theoretical union of $\mathcal{S}_P$ and $\mathcal{S}_Q$ represent the \emph{the logical} 
disjunction ``P or Q''. Last, the logical negation of the proposition P, ``not P'' is represented by all those 
points in $\mathcal S$ that do not lie in $\mathcal{S}_P$. This is, of course, the set-theoretic complement 
$\mathcal{S/S_P}$

This correspondence of the \emph{logical calculus of propositions} about a physical system, on one 
hand and the \emph{algebra} of subsets of the state space on the other, comes naturally and is exactly 
what we have introduced in the third chapter and defined in the previous one, as a \textit{representation}, \ie a map from a 
language to an algebra\footnote{to be more precise this is the special case where we have classical 
rules of logic mapped into subsets, which obey a Boolean algebra}.

So \tb{Sets}, the category of sets, reveals a logic on its own and it is this logic that dictates the behavior 
of propositions about a physical system, which is attached to that category. 

Extending that idea, any topos carries its own logic and is appropriate for representing on its sub-objects 
some language. But now we do not have the restriction of Boolean algebra, as we are dealing with 
topos different that \tb{Sets}
So a representation takes place in a topos and in that way, a formal 
language can be attached to each physical system.

To be more concrete we will start examining the representations first of $\PL S$ and then of that of $\Ln 
S$ languages, in the case of both classical and quantum physics.
.
\begin{section}
{The language $\PL S$}
\label{clasphysPLS}
\end{section}
In chapter \ref{ch5} we introduced the typed language $\PL S$. A representation of that language, in 
general, takes place in a non-Boolean topos. Different theory-types and different physical systems as 
well, invoke different representation in different topoi.

\begin{subsection}
{$\PL S$ in classical physics}
\end{subsection}

In classical physics the topos involved is \tb{Sets}, the category of sets and functions between sets. We recall that in that case: 
\begin{itemize}
  \item The state-space is \emph{symplectic (or Poisson) manifold} $\mathcal{S}$
  \item A physical quantity A is represented by real-valued functions $\breve{A} : \mathcal{S} \rightarrow \mathR$
  \item The representation $\pi_{cl}$ maps the primitive propositions ``A $\varepsilon \Delta$''  to the subset of $
\mathcal{S}$ given by 

\begin{equation}
\label{rep PL(S)}
\pi_{cl} (A~ \varepsilon~ \Delta) := \{{s \in \mathcal{S} | \breve{A}(s) \in \Delta\}} \newline
   \\= \breve{A}^{-1} (\Delta) \\
\end{equation}
\end{itemize}

An extension of the representation to all sentences of $\PL S$ is possible.

Each state \ti{s} assigns to each primitive proposition ``A $\ve~\De$''  a truth value $\nu(A~\ve~\De; s)$, 
which lies in the set \emph{\{false, true\}} (which we identify as \{0,1\}) and is defined as
\begin{equation}
\label{truthvalue}
\nu(A~\ve~\De; s) : = \left\{ 
\begin{array}{l l}
  1 & \quad \mbox{if $\breve{A}(s)$ $\in \De$}\\
  0 & \quad \mbox{if otherwise}\\
\end{array} \right. 
\end{equation}

We see that the truth values, in this case, belong to the set \{0,1\}, which `happens' to be the set of global 
elements $\Ga\O$ of the sub-object classifier \aasdf. This is not an accident, since we have chosen the topos 
\tb{Sets}, where always \aasdf=\{0,1\}.
Hence $\goth H$ is equal to the Boolean Algebra of all Borel subsets of $\cal S$.\\

\begin{figure}
\begin{center}
\scalebox{2}{\includegraphics[width=2in]{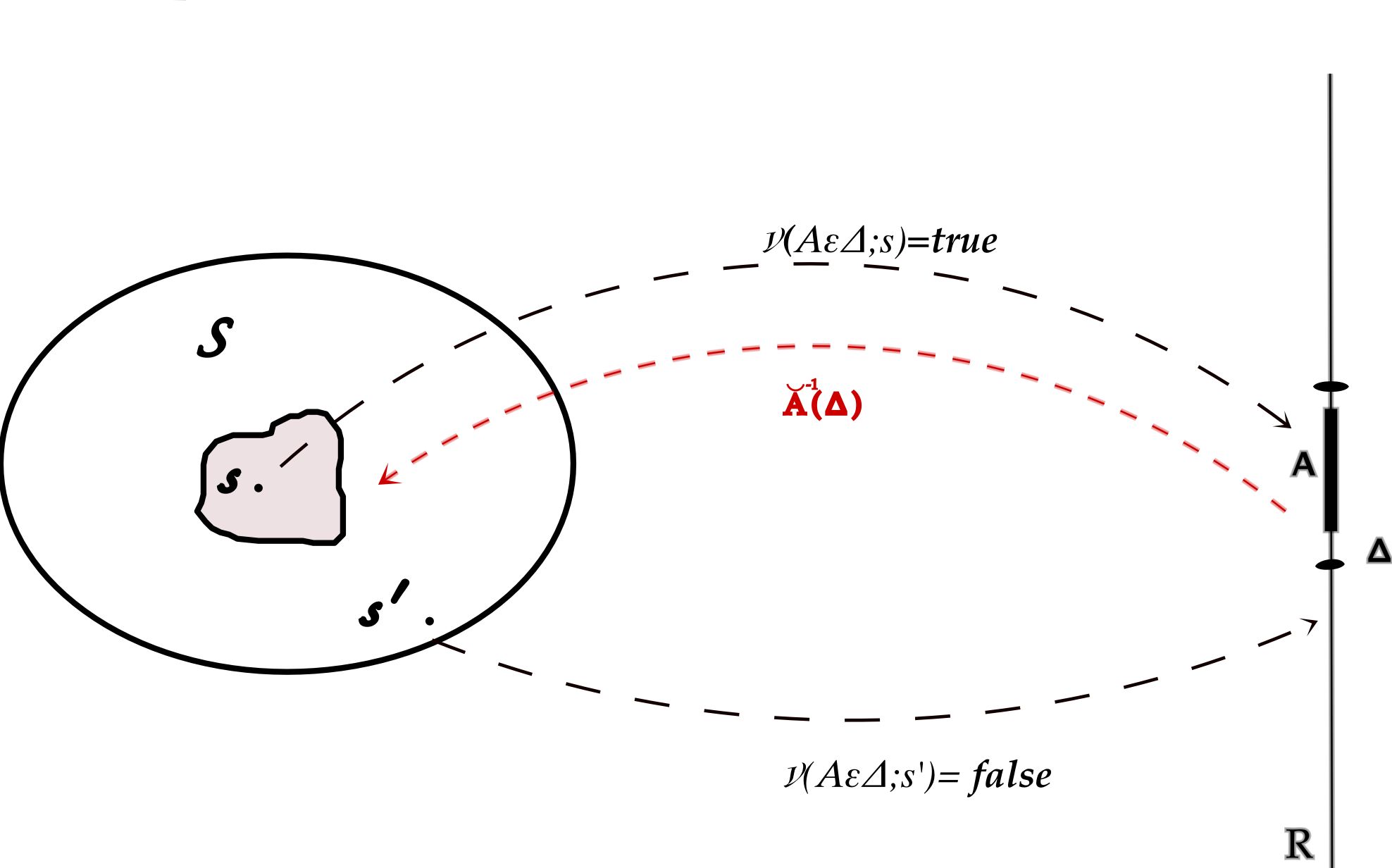}}
\caption{A truth value $\nu$ is assigned to each proposition `${A~\varepsilon~\Delta}$' for each state of the state space}
\label{truediag}
\end{center}
\end{figure}

\subparagraph{For example} we regard, again the Hamiltonian $H = \frac{p^2}{2m} + V(x)$
for the system of the example \ref{example}. The proposition about $S$\\
\begin{equation}
\label{5}
\emph{`The energy H of the system is 5'}
\end{equation}
	of course does not hold true for \emph{all} the states $(p,q)$ of the systems, since we know that 
\emph{H} depends on momentum \emph{p} and position  \emph{q}. So there is a specific set $\tilde{\cal
{S}}$ of pairs $(p,q)$ for which \ref{5} is true, \ie  the equation $H(p,q)=5$ is satisfied for all \emph{p,q} in 
$\va{\cal S}$. The latter is the subset of the state space (symplectic manifold), that the representation $
\pi_{cl}$ maps to (see \ref{rep PL(S)}) \\

Note that we said nothing about the the form of $V(x)$. The details of the Hamiltonian are not 
incorporated in the language, and hence different systems (with different potentials) can have the same 
language. In other words the language $\PL S$ is independent of $V(x)$. This is not true for the 
propositions ``$H~\ve~\De$'';  they depend on the details of the Hamiltonian. This is obvious in our 
example, where the value of the physical quantity `energy' of the system depends on the potential $V(x)
$.

\subsection{$\PL S$ in quantum physics}
In Quantum physics the situation is a quite different. A physical quantity $A$ is represented by a self-adjoint 
operator $\hat{A}$ on a Hilbert space $\Hi$. 

The proposition ``the physical quantity $A$ has a value in the Borel set $\De$'' is now represented by a projection operator\footnote{It is the spectral theorem that guarantees that to each Borel subset $\De$ of the real line $\mathR$ there exists a projection $\hat{E}[A \in \De]$. This projection is interpreted as a proposition.} $\hat{E}[A \in \De]:=\P$ , \ie a self-adjoint operator from $\Hi$ to $\Hi$, such that $\P^2 = \P$.

Hence we have:
\begin{equation}
\label{qrep}
 {\pi (A~\ve~\De) := \E [A~\in~\De]}
\end{equation}
Now the set of all projection operators $\PH$ in $\Hi$ has its own logic; the `quantum logic' of the Hilbert space $
\Hi$. A more thorough study of the quantum logic of projection operator follows in the next chapter.
For the moment it suffices to say that, unlike the distributive lattice of the Heyting algebra (where we represent our language) the logic of the lattice of operators $\P \in \PH$ turns to be non-distributive. 

The quantum logic is incompatible with the intuitionistic logic of $\PL S$ and hence the procedure, followed in the classical case, fails completely. As a consequence we have to modify and `enrich' our scheme in order to construct a viable representation of quantum physics in a topos. As long as we intend to restore a realist interpretation of quantum physics, a different approach, and new mathematical tools are needed. We will present these in the following chapter.

\section{The local language $\cal{L}(S)$: A higher-order language}
\sectionmark{The local language $\Ln S$}
We now introduce and study a language different than $\PL S$. The language presented here, called `local' language $\Ln{S}$ (Bell (\cite{bell}), 
is associated, in a unique way, with each physical system $\S$.

The motivation, as mentioned before, is to construct a topos other than \tb{Sets}, where descriptions of 
theories, such as quantum theory, could be viable. To do so, we no longer restrict ourselves to the 
case where the value space $\mathR$ is fixed, as in classical physics\footnote{In fact this motivation 
comes from Kochen-Specker theorem. In general, it demonstrates that, in a Hilbert space $\Hi$, with dim{$\Hi$}$>$2, it is  impossible to ascribe to an individual quantum system a definite value for each of a set of observables not all of which necessarily commute.}.  The value- (or target)-object $\R_S$ is now 
topos-dependent, and therefore part of the representation. 

To construct a system-dependent quantity-value object we add a symbol $\R$ in the language, to act as 
a linguistic precursor of $\R_S$. In the same way we let the state-object $\Sigma$ be system-dependent 
and hence we add one more symbol `$\Sigma$' to act as the linguistic precursor in the language. But 
then follows that symbols `A: $\Si \rightarrow \O$' should be added too, in order to be represented by an 
arrow in the topos. Finally, we need a symbol `$\O$' as the linguistic precursor of the sub-object 
classifier in the topos. 
We will now, formally, develop the language $\Ln S$.

\paragraph{The symbols of $\Ln S$.} We will present some (the more important ones) from the minimal set of symbols that we need for elementary physics (the whole list of symbols can be found in \cite{AD & CJI II}, p. 17 and onwards). 

\begin{enumerate}
  \item \begin{enumerate}
  \item Basic \emph{type symbols}: 1, $\O$, $\Si$, $\cal R$. The last two are known as ground type 
symbols. They are the linguistic precursors of the state-object and the quantity-value object, respectively.
 
  \item If T is a type symbol then so is PT.
\end{enumerate}
  \item For each type symbol, T, there is associated a countable set of \emph{variables of type T}
  \item  To each pair $(T_1, T_2)$  of type symbols there is associated a set  $F_{(\Ln S )}(T_1,T_2)$ of 
\emph{function symbols}. Such a symbol, A, is said to have \emph{signature} $T_1 \rightarrow T_2$; this 
is indicated by writing A: $T_1 \rightarrow T_2$.
    Some of these sets of functions symbols may be empty. However, particular importance is 
attached to the set, $F_{(\Ln S)}(\Si,\R)$, of function symbols A: $\Si \rightarrow \R $, and we assume this 
set is non-empty.
 \end{enumerate}

The only parts of our language that depend on the system are the function symbols A: $\Si \ra \R$ and 
$F_{(\Ln S)}(T_1,T_2)$; the `physical quantities' of the system. 

Note that, as in the case of the propositional language $\PL S$, different systems can have the same local language $\Ln S$.

\paragraph{The terms of $\cal L$.} There several terms in $\Ln S$. Again we will not enumerate them all, but only those, which are of particular interest and useful for our further study.

Let  $A$ be a physical quantity in the set $F_{\Ln S}(\Si,\R)$, and therefore
a function symbol of signature $\Si\map\R$. In addition, let
$\va\De$ be a variable (and therefore a term) of type $P\R$; and
let $\va{s}$ be a variable (and therefore a term) of type $\Si$.
Then some terms of particular interest to us are  the following:
\begin{enumerate}
\item $A(\va{s})$ is a term of type $\R$ with a free variable,
$\va{s}$, of type $\Si$.

\item `$A(\va{s})\in\va\De$' is a term of type $\O$ with
free variables (i) $\va{s}$ of type $\Si$; and (ii) $\va{\De}$ of
type $P\R$.

\item $\{\va{s}\mid A(\va{s})\in\va\De\}$ is a term of
type $P\Si$ with a free variable $\va{\De}$ of type $P\R$.
\end{enumerate}
As we shall  see, $\{\va{s}\mid A(\va{s})\in\va\De\}$ and
`$A(\va{s})\in\va\De$' are (closely related) analogues of the
primitive propositions ``$A~\ve~\De$'' in the propositional language
$\PL{S}$. However, there is a crucial difference. In  $\PL{S}$,
the `$\De$' in $A~\ve~\De$ is a specific subset of the external (to
the language) real line $\mathR$. On the other hand, in the local
language $\Ln{S}$, the `$\va\De$' in `$A(\va{s})\in\va\De$' is an
\emph{internal} variable within the language.

\paragraph{Adding axioms to the language.} The axioms added to the language must be represented by 
the arrow $$true: \tb{1}_{\tau_\phi} \ra \O_{\tau_\phi}$$

\subsection{Representation of $\cal{L}(S)$ in a topos}
As in the language $\PL S$ again in the case of $\Ln S$ we are looking for representations $\phi$ of it in 
an appropriate topos $\tau_{\phi}$. This corresponds to constructing a physical theory in $\tau_{\phi}$. The choice of both 
representation and topos depend on the theory-type being used. 

 We now list  the $\tau_{\phi}$-representation of the most significant symbols and terms in our language 
$\Ln S$.
 
\begin{enumerate} 
\item 
\begin{enumerate} \item The ground type symbols $\Si$ and
$\cal R$ are represented by objects $\Si_\phi$ and ${\cal R}_\phi$
in $\tau_\phi$. These are identified physically as the state
object, and quantity-value object, respectively.

   \item The symbol $\O$, is represented by
   $\O_\phi:=\O_{\tau_\phi}$, the sub-object classifier of
the topos $\tau_\phi$.

   \item The  symbol $1$, is represented by
$1_\phi:={1}_{\tau_\phi}$,
   the terminal object in $\tau_\phi$.
\end{enumerate}

\item For each type symbol $PT$, we have $(PT)_\phi:=PT_\phi$, the
power object of the object $T_\phi$ in $\tau_\phi$.

In particular, $(P\Si)_\phi=P\Si_\phi$ and $(P{\R})_\phi
=P\R_\phi$.

\item Each function symbol $A:\Si\map\R$ in $\cal{F_{\Ln S}}$$(\Si, \R)$ (\ie\ each
physical quantity) is represented by an arrow
$A_\phi:\Si_\phi\map{\R}_\phi$ in $\tau_\phi$.

We will generally require the representation to be
\emph{faithful}: \ie\ the map $A\mapsto A_\phi$\ is one-to-one.

\item  A term of type $\O$ of the form
{`$A(\va{s})\in\va\De$'} (which has free variables
$\va{s},\va{\De}$ of type $\Si$ and $P\R$ respectively) is
represented by an arrow $\Val{A(\va{s})\in\va\De}_\phi
:\Si_\phi\times P{\R}_\phi\map \O_{\tau_\phi}$.  

We see that the analogue of the `$\De$' used in the
$\PL{S}$-propositions `$A~\ve~\De$' is played by sub-objects of
$\R_\phi$ (\ie\ global elements of $P\R_\phi$). These objects are, of
 course, representation-dependent.

\item Any axioms that have been added to the language are required
to be represented by the arrow
$true:1_{\tau_\phi}\map\O_{\tau_\phi}$.
\end{enumerate}

\paragraph{Topos-language `correspondence'.} It turns out that, using the `local' language $\Ln S$ one 
can find a bidirectional correspondence between the language and the topos used in the 
representation.  This lies in the fact that \emph{for each topos $\tau$ there is a local language $\Ln {\tau}
$}, whose ground-type symbols are the objects of $\tau$ and whose function symbols are the arrows in $
\tau$. It then follows that a representation of a local language, $\cal{L}$, in $\tau$ is equivalent to a 
`translation' of $\cal L$ in $\Ln {\tau}$.

\subsection{${\Ln S}$ in classical physics}
The representations of both languages $\PL S$ and $\Ln S$, in the case of Classical physics are rather 
easier and simpler to study than the application of theory-type `Quantum Physics'. Again we use the 
topos, \tb{Sets}, say $\tau_\sigma$, which is the same for all systems S and all representations $\si$.  
The ingredients of that representation are:

\begin{enumerate}
\item
\begin{enumerate}
        \item The ground-type symbol $\Si$ is represented by
        a symplectic manifold, $\Si_\si$, that is the
        state-space for the system $S$.

        \item The ground-type symbol $\R$ is represented by the
        real line, \ie\ $\R_\si:=\mathR$.

\end{enumerate}

\item Each function symbol $A:\Si\map\cal R$, and hence each
physical quantity, is represented by a real-valued function,
$A_\si:\Si_\si\map\mathR$, on the state space $\Si_\si$.

\item  The term `{$A(\va{s})\in\va\De$}' of type $\O$ (where
$\va{s}$ and $\va\De$ are free variables of type $\Si$ and $P\R$
respectively) is represented by the function
$\Val{A(\va{s})\in\va\De}_\si:\Si_\si\times P\mathR \map\{0,1\}$
that is defined by
\begin{equation}
 \Val{A(\va{s})\in\va\De}_\si(s,\De)=
        \left\{\begin{array}{ll}
            1 & \mbox{\ if\ $A_\si(s)\in \De$;} \\
            0 & \mbox{\ otherwise.}
         \end{array}
        \right. \label{A(s)intildeDeChainCL}
\end{equation}
for all $(s,\De)\in\Si_\si\times P\mathR$.

\end{enumerate}

To make the above clearer, let's again examine the case of a particle $S_1$ moving in one dimension, treated 
classically. We denote the representation with $\si$. The topos $\tau_{\si}$ is \tb{Sets} and $\Si$ is 
represented by the symplectic manifold $\Si_{\phi}:=T^\ast \mathR$. The primary physical quantities that 
we need are \emph{position x, momentum p} and the \emph{energy} of the system $H$, so $$F_{\Ln {S_1}}(\Si, 
{\R}) = \{x, p, H\}.$$ Of course in the three dimensional case the set of physical quantities becomes  $$F_
{\Ln {S_2}}(\Si, {\R}) = \{x, y, z, p_x, p_y, p_z, H\}$$  furthermore we could add the angular momentum in 
our representation to get $$F_{\Ln {S_3}}(\Si, {\R}) = \{x, y, z, p_x, p_y, p_z,J_x, J_y, J_z, H\}.$$ We note 
that the details of the Hamiltonian are encompassed in the representation (in the topos) of the language. 
The same holds in the quantum case, as well.

\subsection{$\cal{L}(S)$ in quantum physics}

A representation for the theory type `quantum physics' is totally different. Since we cannot use anymore 
the representation in topos \tb{Sets} we move to a more complicated one; the topos of presheaves $\cop
$ over a category $\C$. More specifically, the category $\C$ is the category $\VH$ where the objects are the unital commutative von Neumann subalgebras of the algebra $\cal B(H)$  of all bounded operators on the Hilbert space $\Hi$ and the arrows between two objects stand for inclusion (\ie an arrow, from $V'$ to $V$, is assigned if and only if $V' \subseteq V$). A precise definition of this category will be given later. The type symbol $\Si$ of the language is represented by the object $\underline{\Si}$ which is defined as the spectral presheaf. \\

\section{Between two languages} As a conclusion we should highlight here is that, by introducing the local language $\Ln S$ things turn to fit nicely when looking for representations on a topos. This will become more obvious in the next chapters, were we are looking for appropriate representations for those languages in a `quantum topos'. For the time being let us note that by assigning to each physical system the language $\Ln S$ we have a more powerful way of representing terms, that are of physical interest. For example in the propositional language $\PL S$ the state space $\S$, the quantity-value space $\mathR$, the subsets $\De$ and the physical quantities are all external to the language. On the other hand the local language $\Ln S$ has two `ground-type' symbols $\Si$ and $\R$ and a set of `functional symbols' that can be regarded as the `linguistic precursors' of the state object, the quantity-value object and the physical quantities (\ie arrows between the former objects), respectively. In that way the entities that lie outside the propositional language $\PL S$ are all brought `inside' $\Ln S$.  \\

In the next chapters we focus on the representations of both languages $\PL S$ and $\Ln S$ in various topoi (other than \tb{Sets}). This, as we shall see, gives rise to the `employment' of theory-type `quantum physics'.

\chapter{Quantum Topos I: Daseinisation and the representation of $\PL S$ in the topos of Presheaves}   
\chaptermark{Quantum Topos I}
\section{Introduction}
So far we have shown the connection of logic with topos; in fact every topos carries its own logic. We 
claimed that a typed formal language together with an appropriate topos can be used as a starting point 
for writing theories of physics. A language, which could well be the 
propositional $\PL S$ or the local language $\Ln S$, can be attached to each physical system. A 
representation of it, in a topos gives us the description of that system, in the light of  a theory-type (which 
could be classical or quantum physics or even something different!).

Different theory-types, describing the same system, acquire different topoi. The example of harmonic 
oscillator is quite intuitive. An harmonic oscillator can treated both classically and quantum 
mechanically. The choice of one of the two theory-types depends on the nature of the problem, the 
energy scale and others. However in the topos perspective we can examine the (same) physical system 
both ways: either by using the topos \tb{Sets} or using the category of presheaves $\cop$ over a 
category $\C$. 
As noted before, for a physical system there is \tb{not} a single language. As shown 
already, two different systems, $S$ and $S'$ might share the same language, even when they employ  theory-types. 
Moreover, for a given theory-type, we can use the same topos representation of the language, for 
different systems. However, what do depend on the representation are the details of the system \ie the 
energy or the potential of a system.  In other words, for a given theory-type and hence a topos, different 
representations of the language of a system S correspond to different choices of the potential $V(x)$ in 
the Hamiltonian $H=T(\underline{p}) + V(\underline{x})$. 

So far so good. The case where one wants to employ the `quantum topos' and represent on it a 
language, $\PL S$ or $\Ln S$, is more complicated than the classical case. It is know from previous work  
(Isham and Butterfield \cite{IB I}) that the topos now is the \emph{category of presheaves} $\copv$, over 
the category $\cal{V(H)}$, of unital, abelian von Neumann subalgebras of $\cal B(H)$. $\BH$ is the non-
commutative algebra of all bounded operators on the Hilbert space, $\Hi$, of the quantum system. Since physical quantities are represented by self-adjoint operators, one of the main achievements is to 
find a topos representation for these operators.The propositions about a physical system turns out to be represented by \emph{clopen} sub-objects of - what we will later define as - the spectral presheaf $\underline{\Si}$ of $\copv$.  This is where `daseinisation' (D\"oring and Isham \cite{AD & CJI II}) comes in;  a critical step in our construction. Roughly speaking, daseinisation takes the mathematical objects representing propositions, \ie projectors, as explained before, and maps them to a Heyting algebra, which represents an intuitionistic logic. It is a necessary procedure, needed to represent the propositional language $\PL S$ in the topos of presheaves over the context category $\VH$.

The `generalized' truth values $\nu(A ~\ve~\De; \ket\psi)$ assign truth values to the propositions for a quantum state $\ket\psi$. The latter do not anymore belong to a Boolean algebra, since $\Ga\O_{\phi}$, where propositions $\P$ are mapped to,  forms a Heyting algebra. 
  
This is the general scheme showing how the proposition about a quantum system, the topos $\copv$ and a Heyting algebra $\goth H$ are connected. But before focusing on daseinisation and quantum topos, it is exigent to give some useful categorical and algebraic definitions.

\section{Category of Presheaves}

\begin{remark}
The collection of presheaves over a category $\C$ forms a category $\cop$
\end{remark}

Having originally defined categories as collections of objects with arrows between them, by introducing functors we took a step up the ladder of abstraction to consider categories as objects, with functors as arrows between them. Readers are now invited to fasten their mental-safety belts as we climb even higher, to regard functors themselves as objects!\footnote{this `colorful' and flowery introduction is taken from Goldblatt \cite{Gold}, pg 198}

The claim is that from a category $\C$ and a category $\cal D$ we can construct the category, ${\cal D}^{\C}$, whose objects are the the functors from $\C$ to $\cal D$. Thus we need to define the arrows between those `new' objects. These are nothing more that the \emph{natural transformations}. An intuitive idea of those transformations from the functor F: $\C \ra {\cal D}$ to the functor G: $\C \ra {\cal D}$  comes if we image ourselves trying to superimpose or ``slide'' the F-picture onto the G-picture \ie we use the structure of $\cal D$ to translate the former into the latter. This could be done by assigning to each $\C$-object $a$ an arrow in $\cal D$ from the F-image of $a$ to the G-image of $a$. Denoting this arrow by $\tau_a$ we have $\tau_a : F(a) \ra G(a)$. In order for this process to be ``structure-preserving'' we require that each $\C$-arrow $f: a \ra b$ gives rise to a diagram 


\begin{displaymath}
    \xymatrix{
        a \ar[d]_f  &  F(a) \ar[r]^{\tau_a} \ar[d]_{F(f)} & G(a) \ar[d]^{G(f)} \\
        b  &  F(b) \ar[r]_{\tau_b}       & G(b) 
        }
\end{displaymath}
\begin{center}
\textit{\small{Natural transformations between two presheaves}}
\end{center}

that commutes. Thus $\tau_a$ and $\tau_b$ provide a categorial way of turning the F-picture of $f: a \ra b$ into its G-picture. The arrows $\tau$ are called the \emph{natural transformations} while the arrows $\tau_a$ are called the \emph{components} of $\tau$.

The category of presheaves is the  denoted $\cop$ is the category of functors to the category of \tb{Sets} over a general category $\C$.
\subparagraph{Sub-objects of objets in $\cop$.} A sub-object of an object (presheaf) in $\cop$ is a subpresheaf $\underline S$ of a presheaf $\underline T$ over $\C$ such that (i) for all $C\in\C$ we have $\underline{S}(C)\subseteq \underline{T}(C)$ and (ii) the restriction mappings $\underline{S}(i_{V'V}) : \underline{S}(V) \ra \underline{S}(V)$ are the same as the restriction mappings $\underline{T}(i_{V'V}): \underline{T}(V) \ra \underline{T}(V')$ just applied only to the elements in $\underline{S}(V)$.

\begin{definition} A sieve on an object A of a category $\C$ is a collection S of morphisms $f:B \ra A$ in $
\C$ with the property that if $f: B \ra A$ belongs to S and if $g: C \ra B$ is any morphism with codomain 
B, then $f\circ g: C \ra A$ also belongs to S. 
\end{definition}

\begin{displaymath}
    \xymatrix{
        C \ar[d]_{g} \ar[rd]^{f\circ g}  \\
        {B} \ar[r]_{f}       &      \bf{A} }
\end{displaymath}
\begin{center}
\ti{A sieve on A}
\end{center}

\begin{subparagraph} 
{sub-object classifier \½}
\end{subparagraph}
We can define a presheaf   $\underline{\O} : \C \ra \tb{Sets}$ with the use of sieves, as follows. 

\begin{definition}
The presheaf $\underline{\O}$ is defined: 
\begin{itemize}
  \item On objects $A$ of $\C$: $\underline{\O}_A$  the set of all sieves in A and
  \item On morphisms $f : B \ra A$ : $\underline{\O}(f): \underline{\O}_A \ra \underline{\O}_B$ where 
 \begin{equation}
\label{sieve}
\underline{\O}(f)(S) := \{h : C \ra B | f\circ h \in S\} 
\end{equation}
  for all $S \in \underline{\O}_A$. 
\end{itemize}
\end{definition}

A crucial property here, is that $\underline{\O}_A$ is a Heyting algebra where the unit element
${1_{\underline{\O}_A}}$ in $\underline{\O}_A$ is the principal sieve $\downarrow A$ and the null element ${0_{\underline{\O}_A}}$ is the empty sieve $\empty$.

The presheaf $\underline{\O}$ is the sub-object classifier for the category $\cop$ (see \cite{AD & CJI I}, pg 32).
Furthermore one can easily shows that $\cop$ is a Cartesian closed category. In addition, the fact that there is always a sub-object classifier $\underline{\O}$ justify us to say that: 
\begin{remark}
The category $\cop$ is a topos
\end{remark}

\begin{definition}
The category $\VH$ of unital, abelian subalgebras (category of contexts) is defined as the category which has 
\begin{itemize}
  \item Objects: abelian subalgebras $\VH$ of the algebra $\BH$ of all bounded operators on $\Hi$
  \item Morphisms: $i_{V'V}: V' \ra V,  ~~V',V \in Ob(\VH) ~if~ and~ only~ if~ V'\subseteq V $
\end{itemize}
\end{definition}

\subparagraph{$\copv$.} A Special case of $\cop$ is the topos  $\copv$ of all presheaves over the context category $\VH$. It is the main topos that  concerns us in when finding a representation of the projection operators in a Heyting algebra.

\section{Operators, Propositions and Quantum Logic}
We have shown, so far, how propositions about a system are connected to the state space $S$; the 
space where the states of a physical system `live'. In the classical case the propositions are represented 
in a subset\footnote{More precisely \emph{Lebesgue measurable subset}, see Birkhoff-von Neumann \cite{birkhoff von neumann}} of that space (manifold) and they yield a truth value, according to relation \ref{truthvalue}. 
Conversely every subset of $S$ corresponds to a categorical property of the system. Those propositions, once represented by set-theoretic operations, form a Boolean algebra.

In quantum mechanics things are different, since the whole mathematical background is different. Each 
physical system is associated with a (separable) Hilbert space $\Hi$, the unit vectors of which 
correspond to possible physical states of the system.  The state space is the projective unit sphere $S=S(\Hi)$ of a Hilbert space $\Hi$\footnote{However not all subsets of $S$ correspond to quantum-mechanical properties of the system. The latter corresponds only to subsets of the special form $S\bigcap M$, where $M$ is a closed linear subspace of $\Hi$. In particular only propositions of this form are assigned probabilities.}. Each ``observable'' real-valued random quantity is 
represented by a self-adjoint operator A on $\Hi$, the spectrum of which is the set of possible values of 
A. If u is a unit vector in the domain of A, representing a state, then the expected value of the observable 
represented by A in this state is given by the inner product $<Au,u>$. The observables represented by 
two operators A and B are commensurable if and only if A and B commute, $\ie$ AB=BA, \cite{sep}.

\subparagraph{Logic of Quantum Propositions} We should seek a bit deeper to find the `right' representation of propositions in the quantum scheme. The observables that should be regarded as the `encoding propositions' ones are those whose spectrum is just the set \{0,1\}. The operators that yield eigenvalues only in that two-point set are not more nor less than the projection operators $P$, for which $P^2=P$. Projections map the whole (vector) space to a subspace, which, for a projection $P$, we denote by $codP$=`codomain of $P$', and leave the points in that subspace unchanged. This means that we can characterize every subspace $D$ of $\Hi$, using the outcomes \{0,1\} that the set ${\cal P}_D$ of projectors $P$, which project on $D$, gives for every operator in $\Hi$.
Hence \emph{such operators are in one-to-one correspondence with the closed subspaces of the Hilbert space $\Hi$, in the same way that the characteristic functions, in a general topos $\cal E$, are in one-to-one correspondence to the sub-objects of an object in the topos.} In other words, the codomain of a projector $P$ is closed and any closed subspace is the codomain of a unique projection. \\

In such a way we establish a ``\emph{relation between the properties of a physical system on the one hand and projections on the other,(which) makes possible a sort of logical calculus with these}'' (von Neumman \cite{von neumann}, p.253)\\

That relation is quite important since it allows us to examine the propositions in the `logic calculus' framework. 

To see that let us first note that the closed subspaces of $\Hi$ form a poset, if ordered by set-inclusion $\subseteq$.

 \begin{displaymath}
    \xymatrix{
                            & {\{x,y,z\}}                                       &     \\
          {\{x,y\}} \ar[ru]              & {\{x,z\}}   \ar[u]        &  {\{y,z\}}  \ar[lu] \\
          {\{x\}} \ar[u] \ar[ru]              & {\{y\}} \ar[ru] \ar[lu]          &   {\{z\}} \ar[u] \ar[lu]  \\
                           &  {\emptyset}  \ar[ru] \ar[lu]   \ar[u]
         }
\end{displaymath}
\begin{center}
\textit{\small{Ordered by set-incusion, the sub-sets of a set form a poset}}
\end{center}

It turns out that this poset is furthermore a complete lattice, in which the minimum (greatest lower bound or meet) $\bigwedge$ of a set of subspaces is their intersection, while their maximum (least upper bound or join) $\bigvee$ is the closed span of their union. The global minimum of the lattice of the closed subspaces is the null subspace 0, corresponding to the null projection $\hat 0$ and it is contained in every subspace. The global maximum, not surprisingly, is the whole Hilbert space $\Hi$, corresponding to the identity operator $\hat 1$. 
Since a typical closed subspace has infinitely many complementary closed subspaces, this lattice is not distributive; however, it is orthocomplemented.

In view of the above-mentioned one-to-one correspondence between closed subspaces and projections, we may impose upon the set $\cal{L(H)}$ of the projection operators, the structure of a complete orthocomplemented lattice, defining $P\preceq Q$, if and only if $cod(P) \subseteq cod(Q)$ and $P':= \neg P = 1-P$, where we denote $cod(P)$ the (closed) space that $P$ projects to. It is straightforward that $P\preceq Q$ just in case $PQ = QP = P$. This simply says that whenever $P$ projects onto a set $cod(P)$ that is subset of $cod(Q)$, the resulting operator will lie onto $cod(P)$ regardless the order of $P$ and $Q$ acting. 
More generally, if $PQ = QP$, then $PQ = P\wedge Q$, the meet of P and Q in $\cal{L(H)}$, \ie the codomain is just the set-theoretical disjunction of $cod(P)$ and $cod(Q)$.  Also in this case their join is given by $P\vee Q$ = P+Q-PQ \ie the set-theoretical conjunction.

 This construction admits that whenever two projectors $P$ and $Q$ commute the algebra that they form is Boolean. Alternatively the expressions $$PQ=QP$$ and $$P, Q~ lie~in~ a ~common~sub-ortholattice~of~\cal{L(H)}$$ are equivalent.
  These operators are members of a Boolean `block' of $\cal L(H)$.
From a more physical point of view, that brings us to the familiar statement that commuting observables are simultaneously measurable. 

The above relation between the propositions about a quantum system and projections on $\Hi$ enables one to establish and develop a quantum logic. Actually it is the principle ingredient of quantum mechanics; once the quantum-logical skeleton $\cal L(H)$ is in place the remaining statistical and dynamical apparatus of quantum mechanics is essentially fixed. In that sense quantum mechanics reduces to quantum logic and its attendant probability theory. But, as shown above, the logical operations apply only to commuting projections, which are identified with simultaneously decidable propositions. In that case the lattice-theoretic meet and join if projections are interpreted as their conjunction and disjunction. Von Neumman and Birkhoff, in 1936, proposed that the above interpretation can be extended to \emph{non commuting projections} as well. Such a is construction is possible, but immediately faces the problem that the lattice $\Ln H$ is not distributive, \ie does \textbf{not} satisfy the relation 
\begin{equation}
\label{distributive}
\begin{gathered}
a\vee (b\wedge c) = (a\vee b)\wedge (a\vee c)\\~~ or~ equivalently~~ \\
 a\wedge (b\vee c) = (a\wedge b)\vee (a\wedge c)
\end{gathered}
\end{equation}
The obstacle now is the impossibility of giving to these `quantum' propositions a truth-functional interpretation. That led von Neumman and Birkhoff to regard the distributive law as not a universally valid one. A deep analysis of that concept is beyond the scope of that work. However we will come back to that problem later.   

To conclude this short introduction to quantum logic we should mention the `non-classical' logic that  emerges is the reflection of the non-commutability of operators that represent observables. Last, the relation, stated above, between propositions about a quantum system and projections in $\Hi$, was strong enough to lead  Mackey to a system of axioms about quantum propositions - \emph{`questions'} for a system,(Mackey \cite{mackey}), which includes the following:
 
\begin{axiom} 
The partially ordered set of all questions in quantum mechanics is isomorphic to the partially ordered set of all closed subspaces of a separable, infinite dimensional Hilbert space.
\end{axiom}


\section{Daseinisation}
\subsection{Approximating projections from the `context' viewpoint}
In the previous section we described how projections (\ie propositions about a quantum system) are in one-to-one correspondence to the closed subspaces of the Hilbert space $\Hi$. A physical quantity is represented by a self-adjoint operator $\A$ in the algebra $\BH$ of all bounded operators on $\Hi$. A proposition `$A~\ve~\De$' is represented by the projection operator $\E[A~\ve~\De]$ in $\BH$\footnote{in fact it is the spectral theorem that gives the connection between the projection operators and the propositions.More specifically, according to it \emph{in each Borel subset $\De$ of the real line, there exists a projection $\hat{E}[A \in \De]$ which can be interpreted as the proposition of ``the physical quantity $\hat{A}$ has a value in the Borel set $\De$''}}, where $\De \subseteq \mathR$ is a Borel subset.

Now we can choose the unital, abelian algebra $\VH$ of bounded operators in $\Hi$, which is of course a subalgebra of $\BH$, and form a category out of it. The category structure is that of partially-ordered set whose objects are the abelian subalgebras, and in which there is an arrow $i_{V'V}: V' \ra V$ with $V', V \in Ob(\VH)$ if and only if $V'\subseteq V$. 
 That category, sometimes called `the category of contexts' is the basis of topos approach to quantum theory.

From the previous analysis of quantum logic it seems that a projection operator is something quite useful for a linguistic approach to the quantum theory. On the other hand the category of contexts, which is our general tool for a quantum topos, does not need to contain the projections, since the latter could well not belong to an abelian subalgebra; some of them not commute. So what can we say for the projections $\P$ from the point of view of $\VH$?

The answer lies in `daseinisation', \ie define the `closest but not smaller than $\P$' projection operator that \emph{does} belong to $\VH$ 

\paragraph{Daseinisation} Define the `smallest' projection operator $\de(\P)_\nu$ in $\cal V$ that is 
greater than or equal to $\P$:
\begin{equation}
\label{daseinisation}
 \de(\P)_{\nu} :=  \bigwedge\{\hat{Q} \in P(V) | \hat{Q} \succeq \P \}
\end{equation}

In this way we can `approximate' $\P$ from the perspective of $V \in Ob(\VH)$.
Also note that if the projection $\P$ does not belong to $\VH$ then $\delta (\P)$ does belong; it is `sent' there by the process of daseinisation. Clearly, if $\P\in \VH~\Rightarrow~\de(\P)_\nu = \P$ 

Let us note that \ref{daseinisation} gives a minimum over a (maybe infinite) family of projections $\hat Q$ in the projection lattice $\cal P(V)$ of an abelian subalgebra V of $\BH$. Since the lattice $\cal{P(V)}$ is complete\footnote{We know that the projection lattices are complete from the theory of von Neumann algebras} $\de (\P)_V$ is well-defined and in particular it is a projection in V.
In general, daseinisation is a mapping $\P \mapsto \de({\P})_\nu$ (initially introduced by de-Groote as the `V-support' of P) $\Rightarrow
$
$$\E[A\in\De]\mapsto \{\de(\E[A\in\De])_\nu| V \in Ob(\VH)\}$$

This is an important step in our scheme. In fact we shall regard the collection  $\{\de(\E[A\in\De])_\nu| V \in Ob(\VH)\}$ of projection operators for each context $V$ as the appropriate representation of a proposition about the quantum system, rather than the projection $\E[A \in \De]$. The connection with topos is that this collection of projectors is a global element of -what we call- the `outer' presheaf.

{\definition The \emph{outer presheaf} $\underline{O}$ is
defined over the category $\cal{V(H)}$ as follows:
\begin{enumerate}
\item[(i)] On objects
$V\in Ob({\VH})$:  $\underline{O}_V:=\PV$

\item[(ii)] On morphisms $i_{V^{\prime}V}:V^{\prime
}\subseteq V:$ The mapping $\underline{O}(i_{V^{\prime} V}):\underline{O}_V
\map\underline{O}_{V^{\prime}}$ is given by
$\underline{O}(i_{V^{\prime}V})(\hat{\alpha}):=\delta(\alpha)_{V^\prime}$ for
all $\hat{\alpha}\in\PV$.
\end{enumerate}
}

With this definition, it is clear that the assignment $V\mapsto
\delta(\hat{P})_{V}$ defines a global element of the presheaf $\underline{O}$.
Indeed, for each context $V$, we have the projector
$\delta({\hat{P})}_V\in\PV=\underline{O}_V$, and if $i_{V^{\prime}V}:V^{\prime
}\subseteq V$, then
\begin{equation}
\delta\big(\delta({\hat{P}})_V\big)_{V^\prime}
=\bigwedge\big\{\hat{Q}\in\mathcal{P}(V^{\prime})\mid
\hat{Q}\succeq \delta(\hat{P}){_V}\big\}=\delta(\hat{P})_{V^\prime}
\end{equation}
and so the elements $\delta(\hat{P})_{V}$, $V\in Ob(\VH)$, are
compatible with the structure of the outer presheaf. Thus we have
a mapping
\begin{eqnarray}
\label{global}
\delta : \PH  &\map&\Ga\underline{O}                        \nonumber\\
\P  &  \mapsto&\{\delta(\P)_{V} \mid {V\in Ob(\VH) \}}
\end{eqnarray}
from the projectors in $\PH$ to the global elements, $\Ga\underline{O}$, of
the outer presheaf. This is the daseinisation map (cf. \ref{daseinisation})\\

 A presheaf that plays a fundamental role in the construction of representation in a quantum topos is the \emph{Spectral Presheaf} and is actually being regarded as the quantum analogue of phase space.
\begin{definition} {Spectral presheaf}
\label{Def_SpectralPresheaf} The \emph{spectral
presheaf}, $\Siu$, is defined as the following functor from
$\VH^{op}$ to $\tb{Sets}$:
\begin{enumerate}
\item On objects $V$:  $\Siu_V$ is the Gel'fand spectrum of the unital, abelian
subalgebra $V$ of $\BH$; \ie\  the set of all multiplicative
linear functionals $\l:V\map\mathC$ such that $\l(\hat 1)=1$.

\item On morphisms $i_{V^{\prime}V}:V^\prime\subseteq V$:
$\Siu(i_{V^{\prime}V}):\Siu_V\map \Siu_{V^\prime}$ is defined by
$\Siu(i_{V^{\prime}V})(\l):= \l|_{V^\prime}$; \ie\ the restriction
of the functional $\l:V\map\mathC$ to the subalgebra
$V^\prime\subseteq V$.
\end{enumerate}
\end{definition}

It turns out that $\Siu_{\nu}$ is a compact, Hausdorff topological space.\\

\begin{definition} {A Sub-object $\Su$ of the spectral presheaf $\Siu$ is a functor $\Su: {\VH}^{op}\ra \tb
{Sets}$ such that:
\begin{enumerate}
  \item $\Su_{\nu}$ is a subset of $\Siu_{\nu}$ for all $\nu$
  \item If $\nu' \subseteq \nu$ then $\Su(i_{\nu'\nu}) : \Su_{\nu} \ra \Su_{\nu'}$  is just the restriction $
\lambda \mapsto \lambda|_{\nu'}$ (\ie the same as for $\Siu$), applied to elements $\lambda \in \Su_
{\nu} \subseteq \Siu_{\nu}$
\end{enumerate}
}
\end{definition}

In other words, clopen sub-objects $\underline{S}$ of the spectral presheaf $\underline{\Si}$ are sub-objects of $\Siu$, such that for all $\nu$ the 
set $\underline{S_{\nu}}$ is a clopen subset of $\underline{\Si}_{\nu}$

According to the Gel'fand spectral theory, that motivates us to establish a link between the projection operators and sub-objects of the spectral presheaf in the quantum topos. More precisely a projection operator $\hat{a} \in \PV$ correspond to a unique clopen subset of the Gel'fand spectrum, $\Siu_V$. Extending that in the topos framework and with the help of  daseinisation we can formulate the aforementioned connection with topos.

\subsection{From projections to the spectral presheaf} So far we have introduced functors and  presheaves, as a special case of the former; they are just functors which maps from a general category ${\C}^{op}$ to the category of sets, \tb{Sets}. The spectral presheaf is a special object of $\copv$; it maps the objects `subalgebras' to the Gel'fand spectra, which are objects in \tb{Sets}.

To construct a quantum theory of physics we should seek representations of a formal language in the 
topos $\copv$.

Our task is to find the map from the lattice of projectors $\PL S$ to the Heyting algebra $Sub_{cl}(\Siu)$ of clopen sub-objects of the spectral presheaf $\Siu$ 
\begin{equation}
\label{qtmap}
\pi_{cl}: {\PL S}_0 \ra Sub_{cl}{\underline{\Si}}
\end{equation}

where we denote with $\scl$ the set of all clopen sub-objects of $\Siu$.

With daseinisation, a projector $\P$ is transformed to a clopen sub-object $\de(\P)$ of the spectral presheaf in the topos 
$\copv$.

A projection operator $\hat{P} \in \PV$ corresponds to a subset 
\begin{equation}
\label{clopen subset}
S_{\hat P} := \{\lambda \in \Siu_V | \lambda(\hat{P})=1\}
\end{equation} of the Gel'fand spectrum $\Siu_V$ of $V$. This subset turns to be clopen and this the main reason why we need clopen subobjects of the spectral presheaf, as the codomain of the above mapping. Conversely, to each clopen subset $S \subseteq \Siu_V$, there corresponds a unique projection $\hat{P} \in \PV$.

\begin{theorem}
\label{clopen sub-objects}
For each projection operator $\P \in \PH$, the collection of subsets $$S_{\P}:=\{ S_{\delta (\P)_V} \subseteq \Siu_V | V \in Ob(\VH) \}$$
forms a clopen sub-object of the spectral presheaf $\Siu$.
\end{theorem}

The proof can be found in D\"oring and Isham, \cite{AD & CJI I}.

\subparagraph{To summarize the daseinisation process} we can say that two steps are involved in the daseinisation of projectors and the connection of it to an algebra;
\begin{enumerate}
  \item the formula \ref{daseinisation} which `adapts' $\hat P$ to all subalgebras $V$ that do not contain $\hat P$ by approximation from above, and 
  \item for each $\de(\hat{P})_V,~V\in\VH$, construct the corresponding clopen subset of $\Siu_V$ by using \ref{clopen subset}, so that for each abelian subalgebra $V$ we have one clopen subset. 
  \end{enumerate}
Then theorem \ref{clopen sub-objects} assures that the collection of those subsets forms a sub-object of the spectral presheaf $\Siu$.\\

Moreover, what the mapping 
\begin{equation}
\label{clopen}
\de : \PH \ra Sub_{cl}(\Siu)
\end{equation}
 actually does, is to send a projection operator on $\Hi$ to a clopen sub-object of $\Siu$.\\
As explained before the propositions about a quantum system must be represented by a projection operator $\P$. Thus $\delta$ maps a proposition about the system to sub-objects of the spectral presheaf and `restores' the analogy to the situation in classical physics where propositions about the classical system are represented by subsets of the state space. \\

\subsection{The Heyting algebra of the sub-objects of \underline{$\Sigma$}}. We can see why the map \ref{qtmap}, we were looking for, is defined as:
\begin{equation}
\label{qtmap2}
\pi_{qt}(A~ \ve~ \De) := \de(\E[A~ \in~ \De])
\end{equation}
However, that mapping refers to single propositions or `letters' in our language. To extend this definition from ${\PL S}_0$ to $\PL S$, it is necessary to consider the representation of compound propositions, like $``A_1~\ve~\De_1" \vee ``A_2 ~\ve~ \De_2"$. But this \emph{propositional calculus}, that comes up here, must be represented by a specific algebra. 

To examine that algebra, we recall some standard results from general topos theory. Let  $\tau$ be a topos and $d$ an object of it.  The collection $Sub(d)$ of all sub-objects of the object $d$, together with the `object-inclusion' relation\footnote{a monic arrow $f: a \rightarrowtail d$} $(Sub(d), \subseteq)$ is a bounded \emph{lattice}\footnote{see also appendix} with a maximum $1_d$ and a minimum element $0_d$. Furthermore in $Sub(d)$ there is  a \emph{relative pseudo-complement} $-f$, of the monic arrow f, such that $-f: -a \rightarrowtail d$, for every $f: a \rightarrowtail d$, where $-a$ is the greatest element in the lattice $Sub(d)$, disjoint from $a$ (for proof see Goldblatt, \cite{Gold}, p. 180). So $Sub(d)$ is a $relatively~pseudo-complemented$ lattice with a zero element. But this is exactly the definition of a Heyting algebra. So, we conclude that:
$$the~collection~Sub(d),~of~all~sub\mbox{-}objects~of~d~is~a~Heyting~algebra$$

When representing quantum mechanics in a topos, we are interested in the projection operators, which `convey' the propositions, about the quantum system. From the Gel'fand theory it turns out that the topology of the Gel'fand Spectrum of $\Siu_V$ is \emph{extremely disconnected};  the subsets of the Gel'fand spectrum $\Siu_V$ of $V$, that projection operators $\P \in \PV$ correspond to, are \emph{clopen}. An arbitrary family of such clopen subsets of $\Siu_V$ forms a lattice, denoted by ${\mathcal CL}(\Siu_V)$. Moreover, the latter is isomorphic to the lattice $\PV$, $\ie$ the mapping $$\PV \leftrightarrow {\mathcal CL}(\Siu_V)$$ is bijective.
Thus it is natural to ask what happens to the collection $\scl$ of clopen sub-objects of the spectral presheaf $\Siu$. The following theorem gives us the answer.

\begin{theorem}
The collection, $Sub_{cl}(\Siu)$, of all clopen sub-objects of $\Siu$ is a Heyting algebra. 
\end{theorem} 
 
\begin{small} for proof see \cite{AD & CJI II}, pg 13.  \end{small} \\

 This is a very important result. With the operation of  `daseinisation'  we were able to represent the propositions of $\PL S$ about the quantum system in sub-objects of $\Siu$ in topos $\copv$. In other words, we mapped the quantum logic of projection operators into a intuitionistic logic, represented by the Heyting algebra $\goth{H}$ of sub-objects of $\Siu$. There is a subtlety here, worth noting. While the projection operators $\P$ define a lattice in $\Hi$ which is not-distributive, a Heyting algebra $\goth H$ is always distributive. In our case, this lies in the daseinisation procedure; a daseinised bounded operator belongs in the abelian sub-algebra $\VH$ of $\BH$ and commutative operators define a distributive lattice. However, that mapping, from a non-distributive lattice to a distributive one, even it preserves the partial ordering it does not preserve the negation and meet of the lattice, \ie

\begin{gather}
\de(\P \vee \hat{Q}) = \de(\P) \vee \de(\hat{Q}), \\
 \de(\P \wedge \hat{Q}) \preceq \de(\P) \wedge \de(\hat{Q})
\end{gather}
(see \cite{AD & CJI II}, p. 15)\\

Hence the lattice structure  of  $\PV$ is not \emph{completely} preserved, when mapped to $Sub_{cl}(\Siu)$ and this is the `price' we pay, for `distributising' our lattice.
 
\subsection{Sub-objects of the spectral presheaf and global elements of the outer presheaf}
In \ref{global} we defined a relation between the projections $\P$ and the global elements of the outer presheaf $\underline{O}$ \ie a mapping $\delta:~ \PH \ra \Ga\underline{O}$. In \ref{clopen} on the other hand, the relation is between projections on $\Hi$ and sub-objects of $\Siu$. So a sensible question that someone might asks is what is the connection between the global elements $\Ga\underline{O}$ of the outer presheaf and the clopen sub-objects $Sub_{cl}(\Siu)$ of the spectral presheaf.

To answer that, let's first note that the map $$\de : \PH \ra \Ga\underline{O}$$ is injective. Then in $\copv$ there is a monic arrow $\underline{O}\ra P_{cl}\Siu$ and so the map $$\Ga\underline{O} \ra \Ga(P_{cl}\Siu)$$ is injective as well. Keeping in mind that a global element $f: \tb{1} \ra P\Si$ of a power object $P\Si$  is a sub-object $f: S  \rightarrowtail{f} \Si$  of the object $\Si$, it follows from these that the daseinisation map $$\de : \PH \ra \Ga(P_{cl}\Siu)\simeq Sub_{cl}(\Siu). $$ Thus there is an isomorphism between the collection of clopen sub-objects of the spectral presheaf $\Siu$ and the global elements of the power object $P_{cl}\Siu$

\begin{displaymath}
    \xymatrix{
        a \ar[d]_{\delta} \ar[rd]^{\delta}  \\
        \underline{O} \ar[r]_{\name{\iota}}       & P_{cl}\Sigma }
\end{displaymath}
and the above diagram commutes.

\section{Truth values and truth objects}
\subsection{Truth values in Classical Physics}
Once a suitable representation of the propositions, about a system, is achieved, one wants to know whether a certain proposition holds true or not.  
We have discussed shortly the truth values of propositions in classical physics, in section \ref{clasphysPLS}. We gave the general scheme about how a \emph{proposition P} about the system $S$, acquires a \emph{truth value} $\nu$, for every \emph{state} $\sigma$ of the system. 

A proposition, in classical physics, is represented by a subset
 \begin{equation}
\label{ }
\pi_{cl}(A~\ve~\De) := \breve{A}^{-1}(\De) \subseteq \S 
\end{equation}

 of the state space $\S$. Then, the proposition is true in a state $s$ if and only if $s \in \breve{A}^{-1}(\De)$. That idea is captured in fig. \ref{truediag}. This says that a proposition which holds true for a state $s$, might holds false for a state $s'$, and this is not surprising at all; 
\begin{itemize}
  \item physically, the statement that \emph{a proposition about the system, \ti{e.g.} the physical quantity A has a value that lies in the subset $\De \subseteq \mathR$, is satisfied only by the states of the system which `make' that physical quantity of the system acquire such a value}, sounds rather trivial
  \item mathematically, some elements $x$ of $\S$ belong to the subset $K$ of $\S$, while there are other elements $x'$ that do not! 
\end{itemize}

We will see that in quantum mechanics the above scheme does not work so easily.
Thus, it is the state $s$ that assigns a truth value to every primitive proposition ``$A~ \ve~ \De$''. This truth value, denoted by $\nu (A~\ve~ \De;s)$ lies in the set $\{true, false\}$, which we identify with the set $\{0,1\}$.

Now the first (`physical') statement, is summarized in :

\[\nu (A~ \ve~ \De;s) = \left\{ 
\begin{array}{l l} 
  1 & \quad \mbox{if $s \in \pi_{cl}(A~\ve~\De):=\breve{A}^{-1}(\De)$}\\
  0 & \quad \mbox{otherwise}\\
\end{array} \right. \]

while the second (`mathematical') one, can be rewritten as:

\[\nu (x \in K) = \left\{ 
\begin{array}{l l}
  1 & \quad \mbox{if $x$ belongs to K}\\
  0 & \quad \mbox{otherwise}\\
\end{array} \right. \]

The similarity of the above to the characteristic function is not an accident; indeed, the above array defines a `function` that characterizes the elements of the subset $K$ of $\S$. The whole construction fits nicely in the topos \tb{Sets}, of sets; the objects are sets, $K$ is a sub-object of the state object $\S$ and the truth values lie in the sub-object classifier $\O$. In the case of \tb{Sets} the sub-object classifier, of course, is $\O\equiv \{0,1\}$ and hence the sub-object classifier gives us the set of the truth values. 

Now, in a general topos $\tau$, we have:
\begin{displaymath}
    \xymatrix{
        K \ar[r]^{f} \ar[d]_{!} & X \ar[d]^{\chi_{{\tiny K}}} \\
        \bf{1} \ar[r]_{true}       & \Omega }
\end{displaymath}

where $X$ is any object in $\tau$, $f$ is a monic arrow, $\tb 1$ the initial object, $\O$ the sub-object classifier and `!' a unique arrow.  A global element $x$ of $X$, $x \in \Ga X$, is identified with an arrow 
\begin{equation}
\label{ }
\name{x}: \tb{1} \ra X
\end{equation}
The sub-object $K$ of $X$ can also be identified as a global element of the power-object $PX$, of $X$: \begin{equation}
\label{ }
\name{K}: \tb{1} \ra PX
\end{equation}
Then, by definition, we have \begin{equation}
\label{truth value}
\nu (x \in K) := \chi_{K} \circ \name{x} = \chi_K(x)
\end{equation}
 for the truth value of the proposition ``$x \in K$'', where $\chi_{K}$ is the characteristic arrow $\chi_K: X \ra \O$. This is a composition of two arrows; first we pick up a global element of $X$, then map it to the sub-object classifier, to yield a truth value;
 \begin{equation}
\label{globelem}
\chi_K \circ \name{x}: \tb{1} \ra \O
\end{equation}

This can be depicted in the commutative diagram:
\begin{displaymath}
    \xymatrix{
                                              & X \ar[d]^{\chi_K} \\
        \bf{1} \ar[r]_{true}  \ar[ru]^{\name{x}}     & \Omega }
\end{displaymath}

which is the `lower half' of the above, sub-object classifier square. From \ref{globelem} follows that $\nu(x \in K)$ is a global element of $\O$, \ie $$ \nu(x \in K) \in \Ga\O$$.

\subsection{Truth values and truth objects in Quantum Physics}
\subsubsection{Truth values in Quantum Physics} In the quantum case now the previous scheme does not work out and assignment of truth values is rather complicated. The topos used is $\copv$ and all the objects are presheaves. The sub-object classifier is now $\underline{\O} := \O_{\copv}$ at each stage $V$ is the set of all sieves on V, as explained before. The expression \ref{truth value} for a truth value, becomes 
\begin{equation}
\label{quan truth v}
\nu(x \in \underline{K})_V := \{V' \subseteq V | x_{V'} \in \underline{K}_{V'} \}
\end{equation}

for a sub-object $\underline{K}$ of $\underline{X}$, with  global element $\name{x}: \tb{1}\ra \underline{X}$. In a topos framework, the truth values, assigned to propositions, are global elements of the sub-object classifier of $\tau$. Hence they are elements of the Heyting algebra $\Ga \O_{\tau}$. We described that in the classical case, where $\O = \{0,1\}$ and the algebra is Boolean.

Now we want to assign truth values in the propositions about the (quantum) system. As explained before, a proposition is represented in a sub-object $\underline{K} \rightarrowtail \underline{\Si}$  of the state object of $\copv$, $\ie$ the spectral presheaf $\Siu$. Thus we need a global element $\name{s}: \tb{1} \ra \underline{K}$ in \ref{quan truth v}. But from the Kochen-Specker theorem we know that the spectral presheaf cannot have global elements\footnote{more precisely this is the `topos-formulation' of the Kochen-Specker theorem, introduced by Isham and Butterfield \cite{IB I, IB II}}! This is a deeper result in quantum physics; unlike classical physics, in ordinary quantum mechanics, there are no micro-states, \ie~ global elements of the state object, and hence truth values cannot be assigned to propositions.
\subsubsection{The truth object} 
In the absence of global elements of $\Siu$, we have to modify our scheme and refine the truth value assignment in quantum propositions. To overcome that difficulty we have to look deeper in the topos construction and define a truth object. That object will play the role of state; a generalized state which truth values can be assigned to. For that special object we have to add a term in the language $\Ln S$; a linguistic precursor $\mathbb{T}$, of the truth object. This term is of type $PP\Si$ (while $K$ is of type $P\Si$). So, a representation in a topos $\tau_{\phi}$, is denoted by $PP\Si_{\phi}$ ($P\Si_{\phi}$ respectively). Then a global element $\name{\mathbb{T}}: \tb{1}_{\tau_{\phi}} \ra PP\Si_{\phi} $ defines a concrete \emph{truth object} (while a global element $\name{K}: \tb{1}_{\tau_\phi} \ra P\Si_{\phi}$ defines a sub-object of $\Si_{\phi}$), in $\tau_{\phi}$. We recall that, in the topos $\tau_{\phi}$, each proposition is associated to a representation $\Val{ A(\tilde{s}) \in \tilde{\De}}_{\phi}$ of the term $A(\tilde{s}) \in \tilde{\De}$, of type $P\Si$, in $\Ln S$. 

A proposition, which lies in $\name{K}$, will now hold a truth a value:
\begin{equation}
\label{truth value}
\nu(\name{K} \in \mathbb{T}) : \tb{1}_{\tau_{\phi}} \ra \O_{\tau_{\phi}}
\end{equation}

In the case of classical physics, $\O_{\tau_\phi} \simeq \{0,1\}$ , \ref{truth value} becomes 
\[\nu (A~\ve \De; \mathbb{T}) = \left\{ 
\begin{array}{l l}
  1 & \quad \mbox{if $A^{-1}(\De) \in \mathbb{T}$}\\
  0 & \quad \mbox{otherwise}\\
\end{array} \right. \]
which reproduces the previous results. 

To give an example of a truth value of a quantum proposition we will switch back to $\PL S$. In the case of $\copv$, a proposition `$A \in \De$' is represented by a projection $\P$ and $\underline{K}=\de(\P)$. The truth value of that proposition turns to be:
\begin{equation}
\label{q truth value}
\nu(\name{\de(\P)} \in \underline{\mathbb{T}})_V = \{V' \subseteq V | \de(\P)_{V'} \in \underline{\mathbb{T}}_{V'}\}
\end{equation}
where now, we need the restriction to \emph{clopen} sub-objects of $\Siu$. So the truth object $\mathbb{T}$ has to be a global element of $PP_{cl}\Siu$, or -equivalently- a sub-object of $P_{cl}\Siu$ \ie  $$\name{\mathbb{T}} : \tb{1}_{\tau_{\phi}}\ra PP_{cl}\Siu$$. (similarly $\underline{K} \in Sub_{cl}(\Siu)$) 

\subparagraph{In the above brief discussion} we have skipped many (technical) steps.  A more detailed discuss can be found in D\"oring and Isham, \cite{AD & CJI III}, p. 26, 27.
Admittedly this part of C. Isham's and A. D\"oring's work is the most complicated and unfriendly, maybe of all four papers.
However the reader should keep in mind that, since we are not allowed to use global elements of the state object $\Siu$, \ie quantum micro-states, by virtue of Kochen-Specker theorem, we have to generalize the truth value assignment in the topos framework. A construction of a \emph{truth object} is possible, which reproduces the results of the classical case. As can be seen from \ref{truth value}, \ref{q truth value}, the truth object $\mathbb{T}$ plays the role of a state and assigns truth values to the propositions. It is worth noting that, in \ref{truth value}, the truth values lie in the sub-object classifier $\O_{\tau_{\phi}}$, as they should.

\begin{subsection}
{Truth object and global elements of \underline{$\Omega$}}
\label{oo}
\end{subsection}

Let $\psi\in\mathcal{H}$ be a pure state, i.e., a unit vector, and $\widehat
{P}_{\psi}$ a projection operator onto a saubspace of $\mathcal{H}$.

A formal definition of the truth object can be given here, according to the original one, on papers \cite{IB I, IB II}. As described there, to each quantum state $\ket{\psi} \in \Hi$, there corresponds a truth object, $\underline{\mathbb{T}}^{\ket{\psi}}$: 

\begin{equation}
\begin{gathered}
\label{ }
\underline{\mathbb{T}}^{\ket{\psi}} := \{ \widehat{P}_{\psi} \in \underline{O}_V | \mbox{ Prob}(\widehat{P}_{\psi}; \ket{\psi}) = 1\}\\
= \{\widehat{P}_{\psi} \in \underline{O}_V |~ \bra{\psi}\widehat{P}_{\psi} \ket{\psi} = 1 \}
\end{gathered}
\end{equation}
for all stages $V \in Ob(\VH)$. Here the projection operator $\widehat{P}_{\psi}$ represents a proposition about the system, the (normalized) quantum state is $\ket{\psi}$ and Prob$(\widehat{P}_{\psi};\ket{\psi})$ is the probability that the proposition holds true.

Note that the truth object is defined as a sub-object of the outer presheaf, $\underline{O}$. Furthermore, $\underline{O}$ is a sub-object of the presheaf $P_{cl}{\Siu}$ (\ie there is a monic arrow $\underline{O} \ra P_{cl}\Siu$). It follows that $\underline{\mathbb{T}}^{\ket{\psi}}$ is a sub-object of is a sub-object of $P_{cl}\Siu$; 
\begin{equation}
\label{ }
\underline{\mathbb{T}}^{\ket{\psi}} \ra P_{cl}\Siu.
\end{equation}

From the last relation we guess that there must be a way of defining the truth object using the clopen subsets of the spectral presheaf, rather than the outer presheaf. 

To see that, let $\psi\in\mathcal{H}$ be a pure state, i.e., a unit vector, $\widehat
{P}_{\psi}$ the projection onto the corresponding one-dimensional subspace of
$\mathcal{H}$, and let $\operatorname*{Sub}_{cl}(\underline{\Sigma}_{V})$ be
the clopen subsets of $\underline{\Sigma}_{V}$. If $S\in
\operatorname*{Sub}_{cl}(\underline{\Sigma}_{V})$, then $\widehat{P}_{S}
\in\mathcal{P}(V)$ denotes the corresponding projection. The truth object
$\mathbb{T}^{\psi}=(\mathbb{T}_{V}^{\psi})_{V\in\mathcal{V(H)}}$ is given by
\begin{align}
\label{1st}
\forall V\in\mathcal{V(H)}:\mathbb{T}_{V}^{\psi}: &  =\{S\in
\operatorname*{Sub}\nolimits_{cl}(\underline{\Sigma}_{V})\mid\left\langle
\psi\right\vert \widehat{P}_{S}\left\vert \psi\right\rangle =1\}\\
\label{2nd}
&  =\{S\in\operatorname*{Sub}\nolimits_{cl}(\underline{\Sigma}_{V}
)\mid\widehat{P}_{S}\geq\widehat{P}_{\psi}\}\\
\label{3rd}
&  =\{S\in\operatorname*{Sub}\nolimits_{cl}(\underline{\Sigma}_{V}
)\mid\widehat{P}_{S}\geq\delta^{o}(\widehat{P}_{\psi})_{V}\}\\
\label{4th}
&  =\{S\in\operatorname*{Sub}\nolimits_{cl}(\underline{\Sigma}_{V})\mid S\supseteq
S_{\delta^{o}(\widehat{P}_{\psi})_{V}}\}.
\end{align}

\subparagraph{Proof of the equivalence of \ref{1st} - \ref{4th}}
The equivalence $\bra{\psi}\widehat{P}_S\ket{\psi}=1\Leftrightarrow \widehat{P}_S \geq \widehat{P}_{\psi}$, between \ref{1st} and \ref{2nd}, follows from a simple geometrical analysis. The projector $\widehat{P}_S$ projects into a subspace of $\Hi$, say a 2 dimensional space. The term $\widehat{P}_S \ket{\psi}$ gives the projection of $\ket{\psi}$ into that subspace. 

Note that $\ket{\psi}$ is unit vector, $\lVert \ket{\psi} \lVert = 1$ and hence $\lVert \widehat{P}_S\ket{\psi}\lVert\leqslant 1$. Now from $\bra{\psi}\widehat{P}_S\ket{\psi}=1$ it follows that: 
$$\ket{\psi}~//~\widehat{P}_S\ket{\psi}$$ and since $\ket{\psi}$ is a unit vector we get: 
$$\widehat{P}_S\ket{\psi} = \ket{\psi}.$$ 
This says that, whatever the vector $\ket{\psi}$ is, the projection onto $\widehat{P}_S$ is the vector itself.
So the projector$\widehat{P}_S$ leaves the whole projection space $\widehat{P}_{\psi}$ invariant; this only can happen if $\widehat{P}_{\psi}$ is a subset of $\widehat{P}_S$, \ie 
\begin{equation}
\label{ineqproj}
\widehat{P}_S \geq \widehat{P}_{\psi}.
\end{equation}

\begin{figure}[h]
\begin{center}
\scalebox{0.7}{ \includegraphics{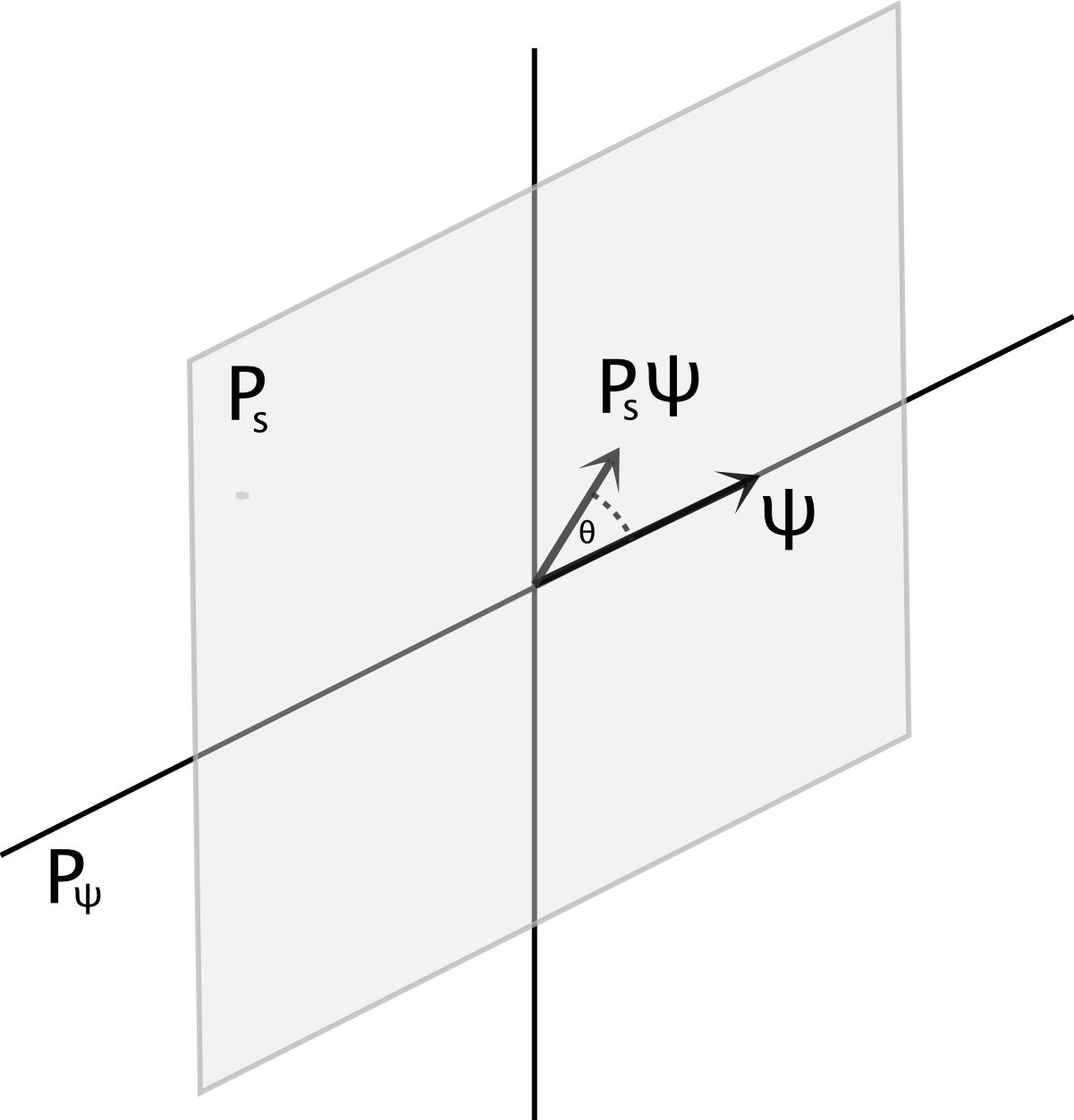} }
\caption{For $\bra{\psi}\widehat{P}_S\ket{\psi}=1$, the angle $\theta$ must be zero.}
\end{center}
\end{figure}

 In a similar way we can show that whenever the inequality \ref{ineqproj} holds, the projection of $\ket{\psi}$ onto $\widehat{P}_S$ yields $\ket{\psi}$, \ie  $\widehat{P}_S\ket{\psi}=\ket{\psi}$ and the equation $\bra{\psi} \widehat{P}_S \ket{\psi}=1$ is true.\\
 
 Last, it is easy to see that the relation \ref{3rd} is true; from the definition of daseinisation we have $\widehat{P}_{\psi} \geq \de^o(\widehat{P}_{\psi})_V$, always (see definition of outer daseinisation, eq. \ref{outerdasein} in the next chapter). And, since $\widehat{P}_S \in \PV$ it follows that $\widehat{P}_S \geq \de^o(\widehat{P}_{\psi})_V$.
  Hence, the corresponding, to the projectors, subsets will follow the `same' inequality, \ie~ $S_{\widehat{P}_S}:=S\supseteq S_{\de^o(\widehat{P}_{\psi})_V}$, as it is in \ref{4th}.\\

\subsubsection{Truth values as global sections of \underline{$\O$}} Our claim here is that the entities $\nu(\name{\underline{S}} \in \mathbb{T}^{\psi})_V$ form a global element of the sub-object classifier $\underline\O$. We will interpret the former as the truth value of a proposition, represented by $\underline{S} \in Sub_{cl}(\Siu)$, about a quantum system, which is in the state$\ket{\psi}$. Obviously, the role of the state of the system, here, is played, again by $\mathbb{T}^{\psi}$ and the truth values do lie in the sub-object classifier.
It is a useful result, which is in accordance to our previous, general, discussion about the truth values.

To see that explicitly, let $\underline{S}\in\operatorname*{Sub}_{cl}(\underline{\Sigma})$ be a clopen
sub-object of the spectral presheaf. Then we have the following:

\begin{lemma}
\label{lemma 1}
$v(\ulcorner\underline{S}\urcorner
\in\mathbb{T}^{\psi})_{V}:=\{V^{\prime}\subseteq V\mid\underline{S}(V^{\prime
})\in\mathbb{T}_{V^{\prime}}^{\psi}\}$ is a sieve on $V$.
\end{lemma}

\begin{lemma}\footnote{This holds, analogously, for sieves on any partially ordered set, not just $\mathcal{V(H)}$}
\label{lemma 2}If $\sigma$ is a sieve on $V\in
\mathcal{V(H)}$ and $V^{\prime}\subset V$, then the pullback $\sigma\circ
i_{V^{\prime}V}$ is given by $\sigma\cap\downarrow\!\!V^{\prime}$.
\end{lemma}

The proofs of those two lemmas can be found in the appendix.

\begin{proposition}
The collection of all sieves $v(\ulcorner\underline{S}\urcorner\in\mathbb{T}^{\psi})_{V}$,
for all $V\in\mathcal{V(H)}$, (see Lemma \ref{lemma 1}) forms a global element of $\underline{\Omega}$.
\end{proposition}

\subparagraph{Proof.}
First let us simplify the notation, a bit, and denote 
\begin{equation}
\begin{gathered}
\label{denote}
\nu(\name{\underline{S}} \in \mathbb{T}^{\psi})_V~:= \check{\sigma}_V\\
\nu(\name{\underline{S}} \in \mathbb{T}^{\psi})_{V'}~:= \check{\sigma}_{V'}
\end{gathered}
\end{equation}
From the definition of the presheaf of sieves $\underline{\O}$ (see \ref{sieve}) we know that 
\begin{equation}
\begin{gathered}
\label{sievepulb}
\underline{\O}(i_{V'V}): \underline{\O}(V) \ra \underline{\O}(V')\\
~~~~~~~~~~~~~~~~~~~~~~ \si \mapsto \si' = \si \circ i_{V'V}
\end{gathered}
\end{equation}
Here, $\si': \tb{1}_{V'} \ra \underline{\O}_{V'}$ and $\si: \tb{1}_{V} \ra \underline{\O}_V$ are global elements of $\underline\O_{V'}$ and $\underline\O_{V}$ respectively, and \ref{sievepulb}
gives a pullback of $\si$ along $i_{V'V}: V' \ra V$, ~\ie
\begin{displaymath}
    \xymatrix{
        \textbf{1}_V=\{\ast\} \ar[r]  \ar[d]_{id} & \underline{\Omega}_V  \ar[d]^{\underline{\Omega}(i_{V'V})} \\
        \textbf{1}_{V'}=\{\ast\} \ar[r]       & \underline\Omega_{V'} }
\end{displaymath}

So, for proving that $\nu(\name{\underline{S}} \in \mathbb{T}^{\psi})_{V} $ form a global element of $\underline{\O}$ we have to show that 
\begin{equation}
\label{ }
\nu(\name{\underline{S}} \in \mathbb{T}^{\psi})_{V'} = \nu(\name{\underline{S}} \in \mathbb{T}^{\psi})_{V} \circ i_{V'V}
\end{equation}
or by using \ref{denote}
\begin{equation}
\label{ }
 \check{\sigma}_{V'} =  \check{\sigma}_V \circ i_{V'V}
\end{equation}
\ie ~$\check{\si}_{V'}$ comes up as a pullback of $\check{\si}_V$.

From Lemma \ref{lemma 2}, it suffices to show that, whenever $V' \subset V$
\begin{equation}
\label{ }
 \check{\sigma}_{V'} =  \check{\sigma}_V \cap\downarrow\!\!V^{\prime} 
\end{equation}

Now  If $V^{\prime\prime}\in \check{\sigma}_{V'} $, then $\underline{S}(V^{\prime\prime})\in\mathbb{T}_{V^{\prime\prime}}^{\psi}$. The last one implies  $V^{\prime\prime}\in \check{\sigma}_V$.

Conversely, if $V^{\prime\prime}\in \downarrow\!\!V^{\prime}$ 
and $V^{\prime\prime}\in \check{\sigma}_V$, then, again, $\underline{S}(V^{\prime\prime}%
)\in\mathbb{T}_{V^{\prime\prime}}^{\psi}$, which implies $V^{\prime\prime}\in \check{\sigma}_{V^{\prime}}$.

We have reached a very important result here; \emph{The collection of all sieves $\nu(\name{\underline{S}} \in \mathbb{T}^{\psi})_{V} $ is a global element of $\underline{\O}$}. The construction of those sieve was possible, only with the help of the \emph{truth object} $\mathbb{T}$, which, as we remarked again, plays the role of the state $\mathbb{T}^{\psi}$, in a quantum system.

\subsubsection{Analogy to Classical Physics; a Neo-Realism is possible} This bring us to a complete analogy to the  classical physics, where, the state space, the state of a system and the truth values of propositions about the system are well defined. Namely, we have constructed, for the case of quantum physics,  something that \emph{ looks like} the state space (\ie the state object $\Siu$), something that \emph{looks like} a state of the system (\ie the truth object $\mathbb{T}$) and something that could well be a truth value of a quantum proposition (\ie the sieves $\nu(\name{\underline{S}} \in \mathbb{T}^{\psi})_V$). Indeed, these sieves \emph{can} be interpreted as truth values; they lie \emph{in} the sub-object classifier $\underline{\O}$ of the topos of presheaves $\copv$ !

This construction looks attractive, but is not so easy to `handle' and comprehend. In classical physics, our topos is \tb{Sets} and the sub-object classifier is just $\O_{\tb{Sets}} \simeq \{0,1\}$. This means that the propositions about the classical system are assigned truth values, which lie in that set; either a proposition (for a particular state of the system) is true, either it is false.

\subsubsection{Partial truth}
In the topos of presheaves $\copv$, even the general scheme remains the same, some things turn out to be, conceptually, quite different. The sub-object classifier now is the presheaf of sieves $\underline{\O}$. So the propositions about the quantum system will acquire truth values, \emph{which lie in $\underline{\O}$}. Let's see how the truth values $\{0,1\}$ can be adapted from the sieves viewpoint. 
We identify the \emph{maximal sieve} as the truth value $\{1\}$ and the \emph{empty sieve} as the truth value $\{0\}$. It turns out that these are \tb{not} the only possible constructions in a collection of sieves. In other words, there is enough space left for truth values that are \tb{neither true nor false!} We call that\footnote{originally in C. Isham's and J. Butterfield's work} state of truth, a \emph{partial truth}.

\begin{displaymath}
    \xymatrix{
       \mbox{maximal sieve} \ar@{.}[dd] \ar[r]  & \mbox{true} \ar@{.}[dd] \\
        \mbox{`mid' sieve} \ar[r]              &   \mbox{partially true} \\
        \mbox{empty sieve} \ar[r] & \mbox{false}}
        \end{displaymath}

Last, to see how a sieve is constructed and the different `grades' of it, we recall that:

\emph{ a sieve on an object $A$ is a collection $S$ of arrows, $f: B \ra A$ with codomain $A$, with the property that if $f: B \ra A$ belongs to $S$ and $g: C \ra B$ is any arrow with codomain $B$, then the composition $f\circ g: C\ra A$ has to belong to $S$.} \\
 
 In other words, the sieve on $A$ is the collection of arrows that `end' to $A$ together with all the `predecessors' of them. Note that, in the definition, we don't need \emph{all} the arrows with codomain $A$. And this allows us to have different sieves on the same object $A$. 
 
 For example, the maximal sieve, by definition, contains all the arrows with codomain $A$, while the empty sieve contains no arrows. A `generic', mid sieve will contains some of those arrows, while others can be excluded;
 
\begin{displaymath}
    \xymatrix{
         \ar[rd]^{f_3}  &                     &               &    \\
         \ar[r]    & V'' \ar[r]_{f_2} \ar@/^/[rr]^{f_{21}} & V' \ar[r]_{f_1} & V \\
         \ar[ru]  & V''' \ar[ru]                &               & }
\end{displaymath}
 we can start collecting the arrows of the sieve $S$, either from $V'$ and backwards, or e.g. from $V''$ an backwards. Each choice gives a different sieve.

\chapter{Quantum Topos II: the representation of $\Ln S$}   
\chaptermark{Quantum Topos II}
\section{Introduction} In the previous chapter we examined a way of representing the propositional language into an appropriate topos; the `quantum topos' $\copv$. 

Let us first recapitulate some of the basic steps, done so far. The main difference (if not the only!) between the two, classical and quantum case, is that the propositions-letters (or symbols) of the language $\PL S$ are now (in the quantum case) projection operators $\P\equiv E[A \in \De]$. We used -as an intermediate step-  that quantum propositions correspond to projection operators $\P$ in $\Hi$; a fact that follows from the spectral theorem\footnote{Actually the identification of propositions `$A \in \De$' with projections $\P = \E[A \in \De]$ is what one does in ordinary Birkhoff-von Neumann quantum logic, and this identification is possible because of the spectral theorem.}. Using  daseinisation we represented the quantum propositions in the, familiar, topos $\copv$, and more precisely, into the clopen sub-objects of the spectral presheaf $\Siu$. The algebra that those sub-objects form turns to be Heyting algebra. So, by putting all these together, we obtained a representation of $\PL S$ - for a system $S$, in the theory-type `quantum-physics' - into a Heyting algebra.

But, so far we have said nothing about physical quantities and moreover the state\footnote{Of course the spectral presheaf $\Siu$ plays the role of the state-object in $\copv$. However this entity was not due to a representation of a linguistic precursor; it shown up as an external object} and the quantity-value objects; the ``space'' that the states of the system live in and the values that physical quantities obtain, respectively.

The skeptical reader will have guessed that we aspire to change the language used. Our intention is to add more (ground-type) symbols to the language and so `upgrade' it into a more powerful one. In other words we want to switch from the propositional language $\PL S$ to a local language $\Ln S$. But we have to keep in mind that this is not a trivial step to do. By adding more and more ground-type symbols in a language one risks its `strength'; too many symbols could weaken it and limit down its applicability. However with the right choice of symbols a much more powerful language is possible. 

So, again we face the problem of `how much information' to put into our language or just let the entities of interest lie in the representation. Secondly, one has to choose an appropriate form of representation, which could well be other than $\cop$. However we keep working in the topos of presheaves $\copv$.
This is actually the problem, discussed again, when we first introduced the local language $\Ln S$. The answer lies in our general conviction that the physical quantities can be of the form
$$A: \Si \ra \R$$ namely represented as an arrow between an object  $\Si$ (identified as the state `space') and an object $\R$ (identified as the quantity-value `space'). 

That scheme, that works so fine in classical physics, is not only abandoned in the quantum case\footnote{The reader should keep in mind that in ordinary quantum-mechanical framework it is the Kochen-Specker theorem that asserts the lack of a, classically conceived, phase space}, but can well serve as a bottom-line for our construction. Hence we are looking for ground-type symbols according to that scheme \ie  linguistic precursors of $\Siu$ and $\R$ objects and $A$ arrows. In fact, our confidence in that scheme, is so strong that will lead us to accept that the `space' of values for the physical quantities is no longer the set of the real numbers! This is not so surprising, since the quantity-value object $\R$ has to be an object in $\copv$ and hence a presheaf itself.

For a symbol $\Si_{\phi}$, that can be construed as the linguistic precursor of the state object $\Si$ we do not have to go far; we just pick up the spectral presheaf $\Siu$ that worked fine in the previous construction; the `success' of it is guaranteed from the fact that the Kochen-Specker theorem can be rephrased in it, in compact way. So the, newly-added, ground-type symbol $\Si_{\phi}$ should be represented in the spectral presheaf $\Siu$.

In a similar way we define ground-type symbols ${\R}_{\phi}$ and $A_{\phi}$ which serve as the linguistic precursors of the quantity value object and the arrows-natural transformations between presheaves, which represent physical quantities.

However it is not easy to work out technically, this simple concept. As we shall see there occur more than one quantity value objects, which are not the set of real numbers. Last, we need not add a symbol $\underline{O}_{\phi}$ in the language $\Ln S$, as a precursor of the outer presheaf $\underline{O}$, since no fundamental physical role is associated to that presheaf.

After that introductory part we are ready to begin with some technical and useful definitions.

\section{An important step}
First we give some definitions in connection with the previous chapter. If $\P$ is a projection operator and $V \in Ob(\VH)$ is a context (or stage) we define

\begin{definition}
The outer daseinisation operation is 
\begin{equation}
\label{outerdasein}
\de^o(\P)_V := \bigwedge \{\hat{Q} \in \PV| \hat{Q} \preceq \hat{P}\}
\end{equation}.
\end{definition}

In a similar way,

\begin{definition}
The inner daseinisation operation is \begin{equation}
\de^i(\P)_V := \bigvee \{\hat{Q} \in \PV| \hat{Q} \succeq \hat{P}\}
\end{equation}
\end{definition}

where `$\preceq, ~\succeq$' denote the usual ordering of projection operators and where $\PV$ is the set of all projection operators in $V$. 

So, with $\de^o(\P)_V$ (respectively $\de^i(\P)$), we can approximate the projection operator $\P$ from above (below), being the smallest (larger) projection in $V$ that is larger (smaller) than or equal to $\P$. The context $V$, in general, does not contain $\P$, but if $\P \in V$ then we get $\de^o(\P)_V = \de^i(\P)_V=\P$.\\

Now with those two definitions we can construct an \emph{outer presheaf} $$\underline{O}: \VH \ra \tb{Sets}$$ and a \emph{inner presheaf} $$\underline{I}: \VH \ra \tb{Sets}$$ (exactly as we did in the previous chapter) as follows
\begin{definition}
The outer presheaf $\underline{O}$, is defined over the category $\VH$ by:
\begin{enumerate}
  \item On objects $V \in Ob({\VH}):~\underline{O}_V~:= \PV$
  \item On morphisms $i_{V'V}:V' \subseteq V$ : define ${\underline O}(i_{V'V}): \underline{O}_V \ra \underline{O}_{V'}$ by $\underline{O}(i_{V'V})(\hat{a}):= \de^o (\hat{a})_{V'}$, for all $a \in \PV$
\end{enumerate}
\end{definition}

\begin{definition}
The inner presheaf $\underline{I}$, is defined over the category $\VH$ by:
\begin{enumerate}
  \item On objects $V \in Ob({\VH}):~\underline{I}_V~:= \PV$
  \item On morphisms $i_{V'V}:V' \subseteq V$ : define ${\underline I}(i_{V'V}): \underline{I}_V \ra \underline{I}_{V'}$ by $\underline{I}(i_{V'V})(\hat{a}):= \de^i (\hat{a})_{V'}$, for all $a \in \PV$
\end{enumerate}
\end{definition}

In words the outer (inner) presheaf maps every subalgebra $V$ to the set of projections $\PV$ of $V$ and assigns an arrow between any two such sets $\underline{O}_V$ and $\underline{O}_{V'},~ V'\ra V$, which is nothing more than the daseinisation $\de^o(\hat{a})_{V'}$ (or $\de^i(\hat{a})_{V'}$ respectively) which adapt the projection $a \in \PV$ to the context $V'$.\\

However, the main part of this chapter is to construct the quantity value presheaf, $\underline{\cal R}$ and show that every physical quantity $A$ can be represented by an arrow between two presheaves in the topos $\copv$, namely a natural transformation $\breve{A}: \Siu \ra \underline{\cal R}$. But before doing that we have to continue with an important step. In the previous chapter we daseinised the projection operators of $\cal{P(V)}$ and this was actually the critical step that allowed us to represent the quantum logic of the lattice $\PV$ to the Heyting algebra of $Sub_{cl}\Siu$. Now and since physical quantities, in Quantum Physics, are being represented by self-adjoint operators it is natural to move on the daseinisation of them.  
\paragraph{Daseinisation of a self-adjoint operator} A formula, about the daseinisation of self-adjoint operators $\A$, that one -naturally- expects to work easily, is to daseinise the spectral projections of $\A$. In other words expand $\A$ into $$\A = \sum_{i=1}^{\infty} a_i  P_i$$ and then daseinise $$\de^o (\A)_V := \sum_{i=1}^{\infty} a_i \de^o (P_i)_V$$
Unfortunately, it turns out that this procedure does not work. On one hand this formula holds only for a \emph{discrete} spectrum and it is quite hard to generalise that in the case of a continuous spectrum. On the other hand, the collection of daseinised projections $\de^o(\P_i)_V, i=1,2\dots$, in general do not form a complete orthonormal set, \ie~ one of the relations 
\begin{equation}
\sum_{i=1}^{\infty} \de^o(\P_i)_V = \hat{1}
\end{equation}
\begin{equation}
\label{daseinised spectral projection}
\de^o(\P_i)_V \de^o(\P_j)_V = \de_{ij} \P_{i}
\end{equation} (where $\delta$ on the right hand side of \ref{daseinised spectral projection} is the usual Kronecker-$\de$  and no summation over $i$-indices is invoked) will not hold true. 

If, instead of the spectral projections, we daseinise the \emph{spectral families} we can overcome the above difficulty. This realization is due to de Groote's work.

A formal definition of \emph{spectrum}, \emph{spectral families} and the \emph{spectral theorem} can be found in the appendix. In words, the spectral theorem associates a spectral family to each self-adjoint operator $\A$ and inversely, for every spectral family it defines a self-adjoint operator.
Now, using that theorem we can get the desired result, namely to daseinise self-adjoint operators.

\begin{definition}
Let $\A$ be an arbitrary self-adjoint operator. Then the outer and inner daseinisations of $\A$ are defined at each stage $V$ as: 
\begin{equation}
 \de^o(\A) := \int_{\mathR} \lambda d(\de^i_V(\hat{E}^A_{\lambda})),
\end{equation} 
\begin{equation}
\de^i(\A) := \int_{\mathR} \lambda d(\bigwedge_{\mu>\lambda} \de^o_V(\hat{E}^A_{\mu})),
\end{equation}respectively
\end{definition}

where $\lambda \ra \hat{E}_{\lambda}$ is a spectral family in $\PH$.\footnote{Note here that for all $\l \in \mathR$ and for all stages $V$ we have \begin{equation*}
\de^i(\E_\l)_V \preceq {\bigwedge_{\mu>\lambda}} \de^o(\E_p)_V
\end{equation*}  and hence, for all $V$, $\de^i(\A)_V  \preceq \de^o(\A)_V$. This explains why the `i' and `o' superscripts in the inner and outer daseinisation are define as they are.}\\

In such a way, admittedly a slightly more technical one, we succeed for the self-adjoint operators what we have already done for the projections; namely an `approximation' for the operator $\A$ from within the contexts $V$, that do not contain $\A$. In the case where $\A \in V$ we get $\de^i(\A)_V=\de^o(\A)_V=\A$, just as in the daseinisation of projections.

Moreover, can shows that the daseinisation procedure (both inner and outer) can be extended to the case that $V$ does not belong to the set $\BH$ of all bounded operators in $\Hi$ or $V$ is not an abelian subalgebra of $\BH$.

\section{The Presheaf \underline{$\mathR^{\succeq}$}}
As pointed in the introduction, our aim is to construct a `quantity-value' presheaf $\underline{\R}$  In other words we need an object (presheaf) $\underline{\R}$ such that an an arrow from the spectral presheaf $\Siu$ to it, will be associated to the physical quantities A, $\ie$ self-adjoint operators. An arrow between any two presheaves cannot be other than a natural transformation and the way we connect it to the self-adjoint operators is not other than the inner and outer daseinisations.

\subsection{The failure of the real-number object}
The arrow corresponding to a self-adjoint operator $\A \in \BH$ is denoted by $\breve{A} : \Siu \ra \underline{\R}$. The mapping we are looking for, is of the form 
\begin{equation}
\label{form }
\breve{A}_V: \Siu_V \ra\underline{ \R_V}
\end{equation}
\begin{equation}
\l \mapsto \breve{A}_V(\l)
\end{equation}

here, $\l \in {\Siu}_V$ is a multiplicative linear functional $$\l \ra \mathC, ~with~ \l(\hat{1})=1,$$  called a \emph{spectral element} of V. So $\l$ must be evaluated on operators lying in V. This is not a problem, since the daseinisations $\de^i(\A)_V$ and $\de^o(\A)_V$ lie in V, even if $\A$ themselves do not. 

Furthermore from the spectral theorem we know that there is a mapping for every self-adjoint operator $\A$\footnote{In case where $\A$ is not self-adjoint, we can always write it as $\A = \A_1 + \imath \A_2$, where $\A_1, \A_2$ are self-adjoint operators. The linearity of the transform $\lambda$ guarantees the viability of the transformation of the non-self-adjoint $\A$.} 
, from the spectral presheaf to the spectrum sp$(\de^o(\A)_V)$ of the operator $\de^o(\A)_V$. This function\footnote{In the original work of D\"oring and Isham, this function is denoted by $\overline{\de^o(\A)_V}$. I prefer to use the `$f$'-symbol to avoid a probable optical confusion with so many lines!}
\begin{equation}
\label{f}
f_{\de^o(\A)_V}: \Siu_V \ra sp(\de^o(\A)_V)
\end{equation}
 represents the self-adjoint operator $\de^o(\A)_V$, of  the commutative von Neumann algebra $V$ and  $f_{\de^o(\A)_V}$ is actually the Gel'fand transform of the daseinisation of $\A$. 
To simplify the notation we denote ${\A}_{\de}$ the (outer) daseinised operator, which lies in $V$, $\ie$ 
\begin{equation}
\label{ }
\A_{\de} := \de^o(\A)_V 
\end{equation}

 Now, \ref{f} becomes: 
\begin{equation}
\begin{gathered}
\label{ }
f_{\A_{\de}}: \Siu_V \ra sp(\A_{\de}) \\
~~~~~ \lambda \mapsto f_{\A_{\de}}(\lambda)
\end{gathered}
\end{equation}

which maps a functional $\lambda : V \ra \mathC$ of $\Siu_V$ to $f_{\A_{\de}}(\lambda)$, which belongs to the spectrum of $\A_{\de}$.

Now this is a strange map, which `transforms a transform'! (the Gel'fand transform). To see that, we note that there is  `trick'  used at that stage:\\
we let the function $f_{\A_{\de}}(\lambda)$ be equal to $\lambda(\A_{\de})$ and hence we get:

\begin{equation}
\begin{gathered}
\label{ }
f_{\A_{\de}} : \Siu_V \ra sp(\A_{\de})\\
\lambda \mapsto \lambda(\A_{\de})
\end{gathered}
\end{equation}

Going further, and since $sp(\A_{\de})$ is a subset of the real numbers $\mathR$, we can rewrite the above relation as:
\begin{equation}
f_{\A_{\de}} : \Siu_V \ra \mathR
\end{equation}
 and so guess that is of the form \ref{form } $\ie$ regard the collection of maps $f_{\A_{\de}}$ as an arrow from $\Siu$ to a presheaf $\underline{\cal R}$. This means that we regard the presheaf $\underline{\cal R}$ (that we are looking for) as the \emph{constant} presheaf $\underline{\mathR}$\footnote{It is known, from general topos theory, that the real-number object, $\underline{\mathR}$ in $\cop$ is the \emph{constant} functor from $\C$ to $\mathR$, $\ie$ every $\C$-object is mapped to the reals.}.   
It turns out that this not the desired formula; the family of Gel'fand transforms, $f_{\A_{\de}}$ of the daseinised operator $\de^o(\A)_V \equiv \A_{\de}, V \in Ob(\VH)$, cannot define an arrow from $\Siu$ to $\underline{\mathR}$. The reader who wishes to see the (technical) justification of that, is prompted to \cite{AD & CJI III}, page 11. 

Summarizing, we see that we cannot employ the constant presheaf as the quantity-value object in the topos of presheaves $\copv$. Thus the object $\underline{\cal R}$ that we are looking for is not the real-number object $\underline{\mathR}$.

At first sight, this inapplicability might seem strange; what we have is a quantum scheme for state-object and physical quantities, which do not hold values on the reals! But, after a second thought we realize that this is not so puzzling. In fact, if we were able to construct a real-number object we would oppose the Kochen-Specker theorem. It this the latter that forbids the assignment of real numbers as values of physical quantities, at least globally.

\subsection{The object \underline{$\mathR^{\succeq}$} and arrows for physical quantities}
We are ready now to shortly introduce the quantity-value presheaf, in $\copv$, which is described in detail in \cite{AD & CJI III} and hence complete the representation of the language $\Ln S$ in the quantum topos $\copv$. We will use the presheaves of `order-reversing' and `order-preserving functions' (see appendix), over the partially order set $\VH$.

We consider the presheaf ${\cal P}^{\succeq}$, where we use the reals $\mathR$\footnote{The set of the reals $\mathR$ is a \emph{totally ordered} set and hence a poset.} for the partially ordered set $\cal P$ and the usual ordering $\leqslant$. 

Now let $\A \in \BH_{sa}$, and let $V \in Ob(\VH)$. Then to each $\lambda \in \Siu_V$ we can associate the function $$ \breve{\de}^o(\A)_V (\lambda) : \downarrow{V} \ra sp(\A)$$
given by  $${\big(}\breve{\de}^o (\A)_V(\lambda){\big)}(V') := \overline{\de^o (\A)_{V'}}{\big(}\Siu(i_{V'V})(\lambda) {\big)}$$
for all $V' \subseteq V$, where $\overline{\de^o(\A)_{V'}}$ is the $f_{\A_\de}$ transform for the daseinisation of $\A$ in the subalgebra $V'$.

Hence $\de^o (\A)_V (\lambda) : \downarrow V \ra sp(\A)$ is an order-preserving function, for each $\l \in \Siu_V$. Since $sp(\A)$ is a a subset of the reals, $sp(\A) \in \mathR$, we can write: $$\breve{\de}^o (\A)_V (\lambda) : \downarrow V \ra \mathR$$

The set of order-reversing functions from $\downarrow V$ to $\mathR$ obtained in this way is now denoted by:
\begin{equation}
\label{natural transformations}
 \breve{\de}^o(\A)_V : \Siu_V \ra \underline{\mathR^{\succeq}}
\end{equation}

We can establish a theorem, which captures our previous claim; the physical quantities are represented by arrows between a state-`space' (the object $\Siu$) and a value `space' \footnote{more accurately: `a candidate for the quantity-value object'} (the presheaf $\underline{\mathR^{\succeq}}$).

\begin{theorem}
The mappings, $\breve{\de}^o(\A)_V, V \in Ob(\VH)$,  are the components of a natural transformation/arrow $\breve{\de}^o(\A) : \Siu \ra \underline{\mathR^{\succeq}}$.
\end{theorem}

The proof of that can be found in D\"oring and Isham, \cite{AD & CJI III}, pg 13.

More precisely, the mapping $\breve{\de}^o(\A)_V$ are arrows from the spectral presheaf $\Siu$ to the object \emph{$\underline{sp(\A)^{\succeq}}_V$}. This object is a sub-object of $\underline{\mathR}^{\succeq}$ and hence that mapping can be also written as $\breve{\de}^o(\A) : \Siu \ra \underline{\mathR^{\succeq}}$.
\newline 

In such a way to each physical quantity $\A$ there is assigned an arrow, from the object $\Siu$ to the object $\underline{\mathR^{\succeq}}$ of $\copv$. That arrow is the natural transformation between the presheaves $\Siu$ and $\underline{\mathR^{\succeq}}$. The technical `vehicle' to succeed that is the daseinisation operation and the Gel'fand transform. Note that this, actually, is a representation\footnote{though it is not the only possible. We can well choose the presheaf $\underline{{\mathR}^{\preceq}}$
or $\underline{{\mathR}^{\leftrightarrow}}$ as the quantity-value objects.}
$A_{\phi} : \Si_{\phi} \ra \R_{\phi}$ in the topos $\copv$, of the `function symbol' $A: \Si \ra \R$ of the language $\Ln S$, that we introduced in the beginning of this chapter.

To define the presheaf $\underline{\mathR}^{\succeq}$ and regard it as a candidate for the quantity-value object,  we used the order-reversing function $\breve{\de}^o (\A)_V(\l): \downarrow V \ra \mathR$. In a similar way we can use the order-preserving functions   $$\breve{\de}^i (\A)_V(\l): \downarrow V \ra \mathR$$ to define a natural transformation $$\breve{\de}^i (\A)_V: \Siu \ra \underline{{\mathR}^{\preceq}}$$ from the spectral presheaf to the presheaf of the real-valued, order-preserving functions on $\downarrow{V}$. The components of this natural transformation are 
$$\breve{\de}^i(\A)_V : \Siu_V \ra \underline{sp(\A)^{\preceq}}_V$$
$$ ~~~~~~~~~~~~~~~~\l \mapsto \breve{\de}^i (\A)_V(\l)$$ 

Combining the functions obtained from both inner and outer daseinisation, now,  we can define the presheaf $\underline{\mathR}^{\leftrightarrow}$, of order-reversing and order-preserving functions.  The functions $$\breve{\de}(\A)_V := \big( \breve{\de}^i(\A)_V(\cdot), \breve{\de}^o(\A)_V(\cdot) \big) : \Siu_V \ra \underline{\mathR}^{\leftrightarrow}_V, ~~V \in Ob(\VH)$$ define an arrow (as components of the natural transformation) $$\breve{\de}(\A) : \Siu \ra \underline{{\mathR}^{\leftrightarrow}}.$$

\section{The Grothendieck completion of the presheaf \underline{$\mathR^{\succeq}$}}

So far we have constructed a general scheme that assigns to each physical quantity an arrow, between two (special) objects, in topos $\copv$. The presheaf $\underline{\cal R}$, which is not the real numbers object, plays the role of  the quantity-value object. The `problem', as we saw, is that we have not one such object, but three; namely the presheaf $\underline{\mathR}^{\succeq}$ of order-reversing, the presheaf $\underline{\mathR}^{\preceq}$ of order-preserving and the presheaf $\underline{\mathR}^{\leftrightarrow}$ of both order-reversing and order-preserving real-valued functions.
Fortunately we are free to choose any of the three presheaves to be the representation $\R_{\phi}$ of the symbol $\R$ of the language $\Ln S$. We choose the presheaf $\underline{\mathR^{\succeq}}$ to be the quantity-value object, \ie ~the representation of $\R$, and hence the arrow $\breve{\de}^o(\A) : \Siu \ra \underline{\mathR^{\succeq}}$ to be the representation of the function symbol $A : \Si \ra \R$.
However all choices are equally good and furthermore there is a way of connecting those different presheaves (more correctly two of them; $\underline{\mathR^{\succeq}}$ and $\underline{\mathR^{\leftrightarrow}}$), according to some topological and algebraic properties of theirs. We will give here the general idea, of how such a connection can be achieved, employing some methods  from algebraic topology.
 
 Such a useful mathematical method, is the \emph{Grothendieck group} of a commutative monoid\footnote{A monoid is a set that is closed under an associative binary operation and has an identity element $e \in S$ such that for all $a \in S$, $e*a=a*e=a$. Unlike a group, its elements need not have inverses. It can also be thought of as a semigroup  with an identity element.}.
   In its simple form it is the universal way of making that monoid\footnote{$S$ might also be an abelian semigroup} $S$ into a group $k(S)$. Some technical details about this, can be found in the appendix. This construction is called \emph{Grothendieck extension} or \emph{Grothendieck completion}.

From the properties of the presheaf $\underline{{\mathR}^{\succeq}}$ follows that it is a \emph{commutative monoid} in the topos $\copv$. Furthermore the collection of global elements  $\Ga\underline{{\mathR}^{\succeq}}$ of $\underline{{\mathR}^{\succeq}}$ has a commutative monoid structure.

On the other hand the set of real numbers $\mathR$ , of standard physics, is an \emph{abelian group} under the ordinary addition of reals\footnote{Furthermore $\mathR$ is a commutative ring.}. Hence it is natural to employ the Grothendieck completion in order to extend the monoid  $\underline{\mathR^{\succeq}}$ to an abelian group. It turns out that this technique can be adapted successfully and give as a result the presheaf  $k(\underline{\mathR^{\succeq}})$. The latter is the `Grothendieck completion' of  $\underline{\mathR^{\succeq}}$; $\ie$ an abelian-group object in the topos $\copv$.

We will skip here the formal definition of $k(\underline{\mathR^{\succeq}})$\footnote{after all this work bristles with definitions!} which can be found in \cite{AD & CJI III} . 

Before going on, let us clarify something. The reason why we want to make $\underline{\mathR^{\succeq}}$ into an abelian group, is not just  the recovery of the similarity of it, to the quantity-value object of the classical physics; namely $\mathR$. It is, rather, the establishment of some relations, necessary for the further development of quantum physics in $\copv$. For example, the relation: 
\begin{equation}
\label{uncertainty}
\nabla (\A):=\breve{\de}^o(\A^2) - \breve{\de}^o(\A)^2
\end{equation}

which can be construed as an `intrinsic dispersion', is not well-defined. Apart from the problematic square $\breve{\de}^o(\A)^2$ of the arrow $\breve{\de}^o(\A) : \Siu \ra \underline{\mathR^{\succeq}}$, nothing assures that we can subtract such arrows. This is related to the fact that the presheaf $\underline{\mathR^{\succeq}}$ is just a monoid and hence the existence of an inverse $-\nu$ of an element $\nu$, where $-\nu, \nu \in \Ga\underline{\mathR^{\succeq}}$, is not guaranteed. 

So, it is the need for a subtraction that compels a group-structure in $ \underline{\mathR^{\succeq}}$, which we achieve through the Grothendieck~k-extension.
 
 As said before the presheaf $k( \underline{\mathR^{\succeq}})$ does exist in $\copv$, as the abelian-group object and, moreover, can be identified as a possible quantity-value object. This means that we can assign to each bounded self-adjoint operator $\A$ an arrow $\left[ \breve{\de}^o(\A) \right] : \Siu \ra k( \underline{\mathR^{\succeq}})$. 
 
 So we now have four candidates of the quantity-value object! Well, things can get simpler, if we regard the relation between the presheaf $\underline{{\mathR}^{\leftrightarrow}}$ of order-preserving and order-reversing real-valued functions and $k( \underline{\mathR^{\succeq}})$. The relation in question is given by the isomorphism 
 \begin{equation}
\label{ }
(\underline{\mathR^{\leftrightarrow}}/\equiv) \cong k( \underline{\mathR^{\succeq}})
\end{equation}
 
 The `$\equiv$' symbol denotes an equivalence relation on $\underline{\mathR^{\leftrightarrow}}_V$, defined by 
 $$ (\mu_1, \nu_1) \equiv (\mu_2, \nu_2) ~\mbox{iff}~ \mu_1 + \nu_1 = \mu_2 + \nu_2$$ for all $V$ and global elements $\mu_1, \mu_2, \nu_1, \nu_2$ of the presheaf $\underline{\mathR^{\leftrightarrow}}$.
 Then $\underline{\mathR^{\leftrightarrow}}/\equiv$ is isomorphic to  $k( \underline{\mathR^{\succeq}})$ under the mapping $$[\mu, \nu] \mapsto [\nu, -\mu] \in  k( \underline{\mathR^{\succeq}})_V$$ for all $V$  and all $[\mu, \nu] \in (\underline{\mathR^{\leftrightarrow}}/\equiv)_V$.


\chapter{Conclusions}  
\paragraph{What we learnt?}
In this long piece of work, there were presented many, physical and mathematical, new ideas. This whole `physics-in-topos' structure, however complicated might looks, can provide us with some quite useful concepts about how a physical theory is constructed. By introducing topos, as a novel tool in physics, we were able to construct a general scheme, modeled after the `properties' of classical physics, wide enough to fit more theories of physics; even such a weird theory as quantum physics! In other words we took the scheme of classical physics and reformulated in the `language' of topos. This allowed us to make the appropriate generalization, in the intuitionistic universe of topos, in order to restore a `realistic'  understanding of quantum physics. The price we pay for that is the abandonment of  the real numbers, as the space where the physical quantities acquire their values. However this is not a discouraging result; from one hand there is not a (well defined and clear) restriction to use the real or complex numbers, as the quantity value space. From the other hand numerous times the use of a non-continuous space-time background has proposed, in order to face the difficulties of quantum gravity. Hopefully, the topos framework is wide enough to fit in a quantum theory of gravity.

Finally and in the writer's point of view, these are the main reasons why the work of C. Isham and A. D\"oring, is remarkable; it proposes novel ideas in \emph{both} fields: conceptual and mathematical-technical.

\paragraph{Open problems and goals} The work presented here is the first part of a project which focuses on establishing a new way of writing theories of physics. So far,  some features of the quantum theory, such as projection operators, self-adjoint operators and operator spectra, have been described in the topos formulation. Though there are still many aspects of quantum theory that have to be incorporated in the general construction. Some of them are the description of commutators within the topos $\copv$, a topos formulation of the uncertainty relations, the superposition of states, composite systems, entaglement and an ongoing discussion on internal or external language formulations.

\paragraph{}
\subsubsection{}
\paragraph{}
\subsubsection{}
\paragraph{}
\subsubsection{}
\paragraph{}
\subsubsection{}
\paragraph{}
\subsubsection{}
\paragraph{}
\subsubsection{}

\subsubsection{Acknowledgments} 
I would like to thank my supervisor, Andreas D\"oring, for his patience in teaching me and introducing to me all these new concepts. Special thanks to my coffee-maker for keeping me awake during all those long nights!
\chapter{Appendix} 

\section{Axioms for CL and IL axiom systems}
Let $a,b$ and $c$ be arbitrary sentences and the symbols `$\wedge$', `$\vee$', `$\supset$' and `$\sim$' denote conjunction, disjunction, implication and negation, respectively. The following twelve forms
\begin{enumerate} 
  \item $a \supset (a \wedge a) $ 
  \item $(a\wedge b) \supset (b \wedge a )$  
  \item $(a\supset b) \supset ((a\wedge c) \supset (b \wedge c))$ 
  \item $((a \supset b) \wedge (b \supset c)) \supset (a\supset c)$ 
  \item $b \supset (a \supset b)$ 
  \item$(a\wedge(a\supset b))\supset b$ 
  \item $a\supset (a\vee b)$ 
  \item$(a \vee b) \supset (b\vee a)$  
  \item $((a \supset c) \wedge (b \supset c)) \supset ((a \vee b) \supset c)$ 
  \item $\sim a \supset (a \supset b)$
  \item $((a \supset b) \wedge (a \supset \sim b)) \supset \sim a$
  \item $a~ \vee \sim a$
\end{enumerate}

are the axioms for \emph{Classical Logic} (CL) (or more generally: ``\ti{the axioms for CL comprise all sentences that are instances of the above twelve forms}", Goldblatt, \cite{Gold}, p. 131.)

The axioms for \emph{Intuitionistic Logic} (IL) are the forms 1-11 (\ie does not contain $a~\vee \sim a$).

\section{Posets, Lattices and Presheaves}   

\subsection{Pre-ordering and partial ordering}
\begin{definition}
A relation ``$\sqsubseteq$''  is called $pre-order$ on a set $S$ if it satisfies: 
\begin{enumerate}
  \item Reflexivity $a\sqsubseteq a$ for all $a\in S$
  \item Transitivity: $a\sqsubseteq b$ and $b\sqsubseteq c$ implies $a\sqsubseteq c$.
\end{enumerate}
\end{definition}

Apart from that we can define the pre-order from a categorical viewpoint. \\
\emph{A category with the property that between any objects p and q there is at most one arrow $p \ra q$, is called a pre-order} (Goldblatt, \cite{Gold}). If $P$ is the collection of objects of a pre-order category then we may define a binary relation $R$ on $P$, \ie a set $R\subseteq P\times P$, by putting \\$<p,q>$ if and only if there is an arrow $p \ra q$ in the pre-order category. 

Then the  relation $R$ has the following properties: 
\begin{enumerate}
  \item \emph{reflexive}, \ie  for each p we have $<p,p>\in R$, and
  \item \emph{transitive} , \ie whenever $<p,q>\in R$ and $<q,s>\in R$, we have $<p,s>\in R$ 
\end{enumerate}

Thus a pre-order category has a natural pre-ordering relation on its collection of objects.. Conversely if we start simply with a set $P$  that is pre-ordered by a relation $R$ then we can obtain a pre-order category.

\begin{definition}
A pre-order that also satisfies antisymmetry \ie: \\
\begin{center}
if $a\sqsubseteq b$ and $b\sqsubseteq a$ then $a=b$ 
\end{center}  is called a partial order. 
\end{definition}
A poset (or partially ordered set) \tb{P}=$\langle P,\sqsubseteq \rangle$ is a set $P$ equipped with partial-order.

We can easily form a category $\C$ from a poset, if we regard $p\in\C$ as objects and assign an arrow 
between p,q $i_{pq}: p \ra q$ if and only if $p \preceq q$. A simple example, given before, is the category of the subsets of a set, when the subsets are ordered by set-inclusion. That inclusion is (i) reflexive, (ii) transitive and (iii) anti-symmetric and hence is a partial order on that set.

\subsection{Lattices and algebras}{\label{lattices}}

In a pre-order $(P, \sqsubseteq)$, a product $p\times q$ is defined by the properties 
\begin{enumerate}
  \item $p\times q \sqsubseteq p, p\times q\sqsubseteq p$, (\ie $p\times q$ is a ``lower bound'' of p and q)
  \item if $c \sqsubseteq p$ and $c\sqsubseteq q$, then $c\sqsubseteq p\times q$, \ie $p\times q$ is ``greater than'' any other lower bound of p and q.
\end{enumerate}
In other words $p\times q$ is a \emph{greatest lower bound} (g.l.b.) of $p$ and $q$. in a poset the g.l.b. is unique, when it exists, and will be denoted $p \sqcap q$.

In a pre-order $(P, \sqsubseteq)$, $p+q$ is defined by the properties 
\begin{enumerate}
  \item $p\sqsubseteq p+q, q\sqsubseteq p+q$, (\ie $p+q$ is an ``upper bound'' of p and q)
  \item if $p\sqsubseteq c$ and $q\sqsubseteq c$, then $p+q\sqsubseteq c$, \ie p+q is ``less than'' any other upper bound of p and q.
\end{enumerate} 

In other words $p+q$ is a \emph{least upper bound} (l.u.b) of p and q. In a poset the l.u.b is unique when it exists, and will be denoted $p\sqcup q$. 
\begin{definition}
A non-empty poset in which any two elements have a l.u.b and a g.l.b. is called a lattice.
\end{definition}

Categorically a lattice is a skeletal pre-order having a product ($a\times b$) and a co-product ($a+b$) for \emph{any} two of its elements.

\begin{definition}
Let $ L$ be a bounded lattice (with 0 and $ 1$), and $ a\in L$. A complement of $ a$ is an element $ b\in L$ such that

    $ a\land b=0$ and $ a\lor b=1$.

An element in a bounded lattice is complemented if it has a complement. A complemented lattice is a bounded lattice in which every element is complemented.

\end{definition}

\begin{definition}
Let $ L$ be a lattice, and $ a,b\in L$. Then $ a$ is said to be pseudocomplemented relative to $ b$ if the set
$\displaystyle T(a,b):=\lbrace c \in L \mid c \wedge a \le b\rbrace$
has a maximal element.
\end{definition}

The maximal element (necessarily unique) of $ T(a,b)$ is called the pseudocomplement of $ a$ relative to $ b$, and is denoted by $ a\to b$. So, $ a\to b$, if exists, has the following property
$\displaystyle c\wedge a\le b$    iff $\displaystyle c\le a\to b.$
If $ L$ has 0, then the pseudocomplement of $ a$ relative to 0 is the pseudocomplement of $ a$.

An element $ a\in L$ is said to be relatively pseudocomplemented if $ a\to b$ exists for every $ b\in L$. In particular $ a\to a$ exists. Since $ T(a,a)=L$, so $ L$ has a maximal element, or $ 1\in L$.

A lattice $ L$ is said to be relatively pseudocomplemented, or Brouwerian, if every element in $ L$ is relatively pseudocomplemented. Evidently, as we have just shown, every Brouwerian lattice contains $ 1$.

\begin{definition}
A lattice is said to be distributive if it satisifes either (and therefore both) of the distributive laws:

    $$ x \land (y \lor z) = (x \land y) \lor (x \land z)$$
    $$ x \lor (y \land z) = (x \lor y) \land (x \lor z)$$
\end{definition}

\begin{definition}
A Boolean algebra $ \goth B$ is a distributive complemented lattice.
\end{definition}

\begin{definition}
A Heyting algebra $ \goth H$ is a relatively pseudocomplemented lattice with a zero element 0.
\end{definition}
\subsection{Presheaves and sieves}

\begin{definition}
A Presheaf \underline{X} is a function on a poset $\C$ that assigns to each $p\in\C$ a set $\underline{X}
_p$ and to each pair $i_{pq} : p \ra q$ a map $\underline{X}_{pq} : \underline{X}_q \ra \underline{X}_p$ 
such that
\begin{enumerate}
  \item $\X_{pp}: \X_p \ra \X_p$ is the identity map $\id_{\X_p}$ on $\X_p$ and
  \item Whenever $p\preceq q \preceq r$ the composite map $\X  \mapright{\X_{rq}} \X_q \mapright{\X_
{qp}} \X_{p}$ is equal to $\X_r \mapright{\X_{rp}} \X_p$, so that $\X_{rp} = \X_{qp} \circ \X_{rq}$
\end{enumerate}
\end{definition}

Equivalently, a \emph{Presheaf} on a poset $\C$ is a contravariant functor $\X$ from the category $\C$ 
to the category \tb{Sets} of normal sets. 

We  have defined what a sieve is in previous chapter. In the case where $\C$ is a poset, a sieve on $p \in 
\C$ is any subset S of $\C$  such that if $r \in \S$ then 
\begin{enumerate}
  \item $r \preceq q$ and
  \item $r' \in S$ for all $r'\preceq r$.
\end{enumerate}  
In other words a Sieve is lower set in a poset.

\subsection{Functions between posets}

\begin{definition}
Let $(\cal{Q}, \preceq$) and Let $(\cal{P}, \preceq$) be partially ordered sets. A function $$\mu: \cal{Q} \ra \cal{P}$$ is \emph{order-preserving} if $q_1 \preceq q_2$ implies $\mu(q_1)\preceq \mu(q_2)$ for all $q_1, q_2 \in \cal{Q}$. It is \emph{order-reversing} if $q_1 \preceq q_2$ implies $\mu(q_1) \succeq \mu(q_2)$. We denote by $\cal{OP(Q,P)}$ the set of order-preserving functions $\mu: \cal{Q} \ra \cal{P}$, and by $\cal{OR(Q,P)}$ the set of order-reversing functions. 
\end{definition}

Furthermore if $\cal P$ is any poset we can define the following presheaf:

\begin{definition}
The $\cal P$-valued presheaf, $\underline{{\cal P}^{\succeq}}$, of order-reversing functions over $\VH$ is defined as follows:
\begin{itemize}
  \item On objects $ V \in Ob(\VH)$: $$\underline{{\cal P}^{\succeq}}_V := \{ \mu : \downarrow V \ra {\cal{P}}~|~ \mu \in {\cal{OR}}(\downarrow V, {\cal P}) \}$$ where $\downarrow V \subset Ob(\VH)$ is the set of all unital von Neumann subalgebras of V.
  \item On morphisms $i_{V'V}: V' \subseteq V :$ The mapping $\underline{{\cal P}^{\succeq}}(i_{V'V}): \underline{{\cal P}^{\succeq}}_V \ra \underline{{\cal P}^{\succeq}}_{V'}$ is given by $$\underline{{\cal P}^{\succeq}}(i_{V'V})(\mu) := \mu_{|V'}$$ where $\mu_{|V'}$ denotes the restriction of the function $\mu$ to $\downarrow V' \subseteq \downarrow V$.
\end{itemize}
\end{definition}

In words, the $\cal P$-valued presheaf $\underline{{\cal P}^{\succeq}}_V$ is a presheaf (and hence a functor from a category $\C$ to $\tb{Sets}$) that assigns to each $\VH$-object the set of all functions $\mu$ from the object (poset) $\downarrow V$ to the poset $\cal P$ and reverses the order of the ordering.\\

There is an analogous definition of the ${\cal P}$-valued presheaf, $\underline{{\cal P}^{\preceq}}$, of order-preserving functions from $\downarrow{V}$ to $\cal P$. It can be shown that \underline{${\cal P}^{\succeq}$} and \underline{${\cal P}^{\preceq}$} are isomorphic objects in $\copv$.
\\

Furthermore we can define a presheaf that combines both order-preserving and order-reversing functions, as follows:
\begin{definition}
The $\cal P$-valued presheaf, $\underline{{\cal P}^{\leftrightarrow}}$, of order-preserving and order-reversing functions over $\VH$ is the presheaf acting:
\begin{itemize}
  \item On objects $ V \in Ob(\VH)$: $$\underline{{\cal P}_V^{\leftrightarrow}} := \{ (\mu, \nu)~|~\mu \in {\cal OP}(\downarrow{V},{\cal P}), ~\nu \in {\cal OR}(\downarrow{V},{\cal P}) \}, $$ 
  where $\downarrow V \subset Ob(\VH)$ is the set of all unital von Neumann subalgebras of V.
  \item On morphisms $i_{V'V}: V' \subseteq V$ : 
  $$\underline{{\cal P}^{\leftrightarrow}}(i_{V'V}): \underline{{\cal P}^{\leftrightarrow}}_V \ra \underline{{\cal P}^{\leftrightarrow}}_{V'}$$ 
$$~~~~~~~~~~~~~~~~~~~~~(\mu, \nu) \mapsto (\mu|_{V'}, ~\nu|_{V'}) $$
  where $\mu_{|V'}$ denotes the restriction of the function $\mu$ to $\downarrow V' \subseteq \downarrow V$ and analogously for $\nu|_{V'}$.
\end{itemize}
\end{definition}

\section{Borel sets and algebra}   
The Lebesgue measure is an extension of the classical notions of length and area to more complicated sets.\begin{definition}
Given an open set $S\equiv \sum_k(a_k,b_k)$ containing disjoint intervals, the Lebesgue Measure is defined by $$\mu_L (S) := \sum_k (b_k - a_k)$$ Given a closed set $S'\equiv [a,b] - S$   $$\mu_L (S') := (b-a) - \sum_k (b_k - a_k)$$
\end{definition}
For example, a unit line element has Lebesgue measure 1; the Cantor set has Lebesgue measure 0. The Minkowski measure of a bounded closed set is the same as its Lebesgue measure. 

\begin{definition}
A Borel $\sigma$-Algebra on a topological space $X$ is a $\si$-algebra of subsets of X associated to the topology of X. 
\end{definition}

The Borel $\si$-algebra is defined to be the minimal $\si$-algebra containing the open sets. However sometimes is defined as the minimal $\si$-algebra containing the compact sets).

\begin{definition}
A Borel set X is an element of a Borel $\si$-algebra. A Borel subset is a subset of the Borel set X.
\end{definition}
Roughly speaking, Borel sets are the sets that can be constructed from open or closed sets by repeatedly taking countable unions and intersections. Formally, the class B of Borel sets in Euclidean $\mathR^n$ is the smallest collection of sets that includes the open and closed sets such that if $E, E_1, E_2, \ldots$ are in B, then so are $\bigcup_{i=1}^{\infty}E_i, ~\bigcap_{i=1}^{\infty}E_i$ and $\mathR^n \backslash E$, where $F\backslash E$ is the set difference.

Examples of  Borel sets are the set of rational numbers $\mathbb{Q}$, the Cantor set and others.

\section{$C^*$-~ and Von Neumann algebras}    

\begin{definition}
A Banach space $ (X,\lVert \,\cdot\, \rVert )$ is a normed vector space such that $ X$ is complete under the metric induced by the norm $ \lVert \,\cdot\, \rVert $
\end{definition}

Hilbert spaces with their norm given by the inner product are examples of Banach spaces. While a Hilbert space is always a Banach space, the converse need not hold.

\paragraph{Joke 1} If a space $(X,\lVert \,\cdot\, \rVert )$ is normed, complete and \emph{yellow}, then it is  a \emph{Bananach space}.

\begin{definition}
An algebra $\goth u$ (over $\mathR$ or ~$\mathC$) with unit $I$ is said to be a \emph{normed algebra} when $\goth u$ is a normed space such that $\Vert AB\Vert \leqslant \Vert A\Vert \cdot \Vert B \Vert$ for all $A$ and $B$ in $\goth u$ and $\Vert{I} \Vert = 1$. If $\goth u$ is a Banach space relative to this norm, $\goth u$ is said to be a \emph{Banach algebra}.
\end{definition}

\begin{definition}
A $C^*$-algebra $A$ is a complex Banach algebra $\goth u$ (with a unit element $I$)  that $ \Vert a^*a\Vert = \Vert a\Vert ^2$ for all $ a \in \goth{u}$
\end{definition}

Let $\Hi$ be an Hilbert Space, $\BH$ the algebra of bounded operators in $\Hi$ and $ \mathcal{F} \subset \BH$. Then,

\begin{definition}
The commutant of $ \mathcal{F}$, usually denoted $ \mathcal{F}'$, is the subset of $\BH$ consisting of all elements that commute with every element of $ \mathcal{F}$, that is
$ \mathcal{F}'=\{T \in \BH:\; T\cdot S=S\cdot T \;\;\; \forall S \in \mathcal{F}\}$

The double commutant of $ \mathcal{F}$ is just $ (\mathcal{F}')'$ and is usually denoted $ \mathcal{F}''$.
\end{definition}

\begin{definition}
A von Neumann algebra (or $ W^*$-algebra) $ \mathcal M$ is a $ C^*$-subalgebra of $\BH$ that contains the identity operator and satisfies the following  condition:

    $ \mathcal M = \mathcal M''$, i.e. $ \mathcal M$ equals its double commutant.
\end{definition}

Examples:

   1. $\BH$ is itself a von Neumann algebra.

   2. $ L^{\infty}(\mathbb{R})$ as subalgebra of $ B(L^2(\mathbb{R}))$ is a von Neumann algebra.

\section{Spectral Theorem}   
\subsection{Spectrum}
\begin{definition}
If $A$ is an element of a Banach algebra $\goth u$, we say that a complex number $\l$ is a \emph{spectral value} for $A$ (relative to $\goth u$) when $A - \l I$ does not have a two-sided inverse in $\goth u$. The set of spectral values of $A$ is called the \emph{spectrum} of $A$ and denoted by $sp_{\goth u}(A)$.
\end{definition}

\begin{theorem}
 $ \lambda$ is a spectral value of $ A$ if and only if $ \lambda-\mu$ is a spectral value of $ A-\mu I$.
\end{theorem}
\paragraph{Proof}
 Note that
$\displaystyle A-\lambda I=(A-\mu I)-(\lambda I-\mu I) =(A-\mu I)-(\lambda-\mu)I $
and thus $ A-\lambda I$ is invertible if and only if $ (A-\mu I)-(\lambda-\mu)I$ is invertible. Equivalently, $ \lambda$ is a spectral value of $ A$ if and only if $ \lambda-\mu$ is a spectral value of $ (A-\mu I)$, as desired.

\begin{subsection}
{Spectral Projections}
\end{subsection}

Let ${\cal A}_0$ be an abelian von Neumann algebra, generated by a normal operator $A$ acting on a Hilbert space $\Hi$, $S$ a Borel subset of $\mathC$ and $g$ the characteristic function of $S$. Then the projection $E(S)=g(A)$ is a \emph{spectral projection for $A$ corresponding to the Borel subset S of $\mathC$}.

 In the case where the spectrum of $\A$ is discrete we have:
$$\A := \sum_{i=1}^{\infty} a_i \P_i,~~where ~a_i \in sp(\A)$$

\subsection{Spectral families}
\begin{definition}
A spectral family is a family of projection operators $\E_\l , \l \in \mathR$ with the following properties:\\
\begin{enumerate}
  \item If $\l_2\leqslant \l_1$ then $\E_{\l_2} \preceq \E_{\l_1}$
  \item The net $\l \mapsto \E_\l$ of projection operators in the lattice $\PH$ is bounded above by $\hat 1$ and below by $\hat 0$. In fact $$lim_{\l\ra \infty}\E_\l = \hat{1}$$ 
  $$lim_{\l\ra -\infty} \E_{\l}=\hat{0}$$ 
  \item The map $\l \mapsto \E_\l$ is right continuous: $$\bigwedge_{\e | 0} \E_{\l+1}=\E_\l$$ for all $\l \in \mathR$.
\end{enumerate}
\end{definition}

\subsection{Spectral Theorem}
\begin{definition}
Let $\Hi$ be a Hilbert space, $\BH$ the set of bounded linear operators from $\Hi$ to itself, $T$ an 
operator on $\Hi$, and $\si(T)$ the operator spectrum of $T$. Then if $T \in \BH$ and $T$ is normal, 
there exists a unique resolution of the identity $E$ in the Borel Subsets of $\si (T)$ which satisfies 
\begin{equation}
\label{spectral}
T=\int_{\sigma(T)}{\lambda dE(\lambda)}
\end{equation}
furthermore, every projection $E(\omega)$ commutes with every $S\in \BH$ that commutes with $T$.
\end{definition}

\section{Gel'fand Transform}                       
The Gelfand transform is a very useful tool in the study of commutative Banach algebras and, particularly, $ C^*$-algebras.

Let $ \mathcal{A}$ be a Banach algebra over $ \mathbb{C}$. Let $\Lambda$ be the space of all multiplicative linear functionals in $ \mathcal{A}$, in the senses:
$$\l(ax+by)=a \l(x)+b \l(y)~ \mbox{  and   }~ \l(xy)=\l(x)\l(y).$$
  Let $ C(\Lambda)$ denote the algebra of complex valued continuous functions in $\Lambda$.
\begin{definition}
The Gelfand transform is the mapping
$$ \widehat{}\;\;:\mathcal{A} \longrightarrow C(\Lambda)$$ $$x \longmapsto \widehat{x}$$
where $ \widehat{x} \in C(\Lambda)$ is defined by $ \;\;\widehat{x} (\l) := \l(x), \;\;\;\forall \l \in \Lambda$
\end{definition}

The Gelfand transform is a continuous homomorphism from $ \mathcal{A}$ to $ C(\Lambda)$ and is automatically bounded.

For example, if $\mathcal{A} = L^1(R)$ with the usual norm, then $\mathcal{A}$ is a Banach algebra under convolution and the Gelfand transform is the Fourier transform.

\section{Clopen sets}                                                  
\begin{definition}                              
A clopen set is a set which is both open and closed
\end{definition}

In any topological space $X$, the empty set and the whole space $X$ are both clopen.

Now consider the space $X$ which consists of the union of the two intervals [0,1] and [2,3]. The topology on $X$ is inherited as the subspace topology from the ordinary topology on the real line $\mathR$. In $X$, the set [0,1] is clopen, as is the set [2,3]. This is a quite typical example: whenever a space is made up of a finite number of disjoint connected components in this way, the components will be clopen.

As a less trivial example, consider the space $\mathbb{Q}$ of all rational numbers with their ordinary topology, and the set $A$ of all positive rational numbers whose square is bigger than 2. Using the fact that $\surd 2$ is not in $\mathbb{Q}$, one can show quite easily that $A$ is a clopen subset of$\mathbb  Q$. (Note also that $A$ is not a clopen subset of the real line $\mathR$; it is neither open nor closed in $\mathR$.)
\section{Kochen-Specker Theorem}                        
Let $\Hi$ be a Hilbert space, $\BH$ the set of all bounded self-adjoint operators $\A$ in $\Hi$. Then

\begin{theorem}
if the dimension of $\Hi$ is greater than 2, then, there, does not exist any valuation function 
$V_{\Psi}: \BH \ra \mathR$ from $\BH$ to the reals such that, the functional composition principle is satisfied for all $\A \in \BH$ and all $\Psi \in \Hi$. 
\end{theorem}

\section{Proofs of lemmas of section \ref{oo}. }   

\subparagraph{Proof of lemma \ref{lemma 1}}
As usual, we identify an inclusion morphism $i_{V^{\prime}V}$ with $V^{\prime}$
 itself, so a sieve on $V$ consists of certain subalgebras of $V$. We have
to show that if $V^{\prime}\in v(\ulcorner\underline{S}\urcorner\in
\mathbb{T}^{\psi})_{V}$ and $V^{\prime\prime}\subset V^{\prime}$, then
$V^{\prime\prime}\in v(\underline{S}\in\mathbb{T}^{\psi})_{V}$. Now,
$V^{\prime}\in v(\ulcorner\underline{S}\urcorner\in\mathbb{T}^{\psi})_{V}$
means that $\underline{S}(V^{\prime})\in\mathbb{T}_{V^{\prime}}^{\psi}$, which
is eqivalent to $\underline{S}(V^{\prime})\supseteq\underline{S}_{\delta
^{o}(\widehat{P}_{\psi})_{V^{\prime}}}$. Here, $\underline{S}_{\delta
^{o}(\widehat{P}_{\psi})_{V^{\prime}}}$ is the component at $V^{\prime}$ of
the sub-object $\underline{S}_{\delta^{o}(\widehat{P}_{\psi})}=(\underline
{S}_{\delta^{o}(\widehat{P}_{\psi})_{V}})_{V\in\mathcal{V(H)}}$ of
$\underline{\Sigma}$ obtained from daseinisation of $\widehat{P}_{\psi}$.  We know that the sub-object $\underline{S}%
_{\delta^{o}(\widehat{P}_{\psi})}$ is optimal in the following sense: when
restricting from $V^{\prime}$ to $V^{\prime\prime}$, we have $\underline
{\Sigma}(i_{V^{\prime\prime}V^{\prime}})(\underline{S}_{\delta^{o}(\widehat
{P}_{\psi})_{V^{\prime}}})=\underline{S}_{\delta^{o}(\widehat{P}_{\psi
})_{V^{\prime\prime}}}$,\footnote{Note the equality here. For a sub-object an inclusion `$\supseteq$' would be enough} i.e., the restriction is surjective. By assumption,
$\underline{S}(V^{\prime})\supseteq\underline{S}_{\delta^{o}(\widehat{P}_{\psi
})_{V^{\prime}}}$, which implies%
\[
\underline{S}(V^{\prime\prime})\supseteq\underline{\Sigma}(i_{V^{\prime\prime
}V^{\prime}})(\underline{S}(V^{\prime}))\supseteq\underline{\Sigma}(i_{V^{\prime
\prime}V^{\prime}})(\underline{S}_{\delta^{o}(\widehat{P}_{\psi})_{V^{\prime}%
}})=\underline{S}_{\delta^{o}(\widehat{P}_{\psi})_{V^{\prime\prime}}}.\footnote{The equality, in the last step of the proof, is important; for if it was an inclusion `$\subseteq$' we would not be able to obtain the desirable result!} 
\]
This shows that $V^{\prime\prime}\in\mathbb{T}_{V^{\prime\prime}}^{\psi}$ and
hence $V^{\prime\prime}\in v(\ulcorner\underline{S}\urcorner\in\mathbb{T}%
^{\psi})_{V}$.

\subparagraph{Proof of lemma \ref{lemma 2}}
By definition, the pullback $\sigma\cdot i_{V^{\prime}V}$ is given by%
\[
\sigma\cdot i_{V^{\prime}V}:=\{i_{V^{\prime\prime}V^{\prime}}\mid i_{V^{\prime}%
V}\circ i_{V^{\prime\prime}V^{\prime}}\in\sigma\}.
\]
We now identify morphisms and subalgebras as usual and obtain (using the fact
that $V^{\prime\prime}\subseteq V^{\prime}$ implies $V^{\prime\prime}\subset
V$)%
\[
\{i_{V^{\prime\prime}V^{\prime}}\mid i_{V^{\prime}V}\circ i_{V^{\prime\prime
}V^{\prime}}\in\sigma\}\simeq\{V^{\prime\prime}\subseteq V^{\prime}%
\mid V^{\prime\prime}\in\sigma\}=\downarrow\!\!V^{\prime}\cap\sigma.
\]

\section{Grothendieck group construction}                            
In its simplest form, the Grothendieck group of a commutative monoid is the universal way of making that monoid into an abelian group.

Let $ S$ be an abelian monoid. The Grothendieck group of $ S$ is $ k(S) = S\times S/\mathord{\sim}$, where $ \sim$ is the equivalence relation: $ [s,t] \sim [u,v]$ if there exists $ r \in S$ such that $ s+v+r = t+u+r$. This is indeed an abelian group with identity element $[s,s]$ (any $ s \in S$) and inverse $ -[s,t] = [t,s]$. It is common to use the suggestive notation $ t-s$ for $[t,s]$.\\

 Now, the Grothendieck group $k(S)$ of a commutative monoid  $S$ has the following property:\\
 There exists a monoid homomorphism

    $$r:S \ra k(S) $$

such that for any monoid homomorphism

    $$ f: S \ra k(T)$$

from the commutative monoid $S$ to an abelian group $k(T)$, there is a unique group homomorphism

   $$ g : k(S) \ra k(T ) $$

such that the diagram 

\begin{displaymath}
    \xymatrix{
        S \ar[r]^{r} \ar[rd]_f & k(S) \ar[d]^{g} \\
        & k(T) }
\end{displaymath}
\ie~ $f=g\circ r$

The Grothendieck group construction, in the language of Category theory, is a functor\footnote{more precisely it is left adjoint to the forgetful functor from the category of abelian groups to the category of commutative monoids.} $r$ from the category of abelian monoids $S$ to the category of abelian groups $k(S)$. A morphism $$ h\colon S \to R$$ induces a morphism $$ k(h)\colon k(S) \to k(R)$$ which sends an element $ (s^+,s^-) \in k(S)$ to $ (h(s^+),h(s^-)) \in k(R)$

and the diagram 

\begin{displaymath}
    \xymatrix{
        S \ar[r]^{r} \ar[d]_h & k(S) \ar[d]^{k(h)} \\
        R \ar[r]_{r'}            & k(R) }
\end{displaymath}
commutes.

\end{document}